\documentclass[twocolumn]{aastex631}

\usepackage{natbib}
\usepackage{graphicx}
\usepackage{hyperref}
\usepackage{amsmath}
\usepackage{lipsum}
\usepackage{grffile}
\usepackage{hyphenat}

\begin{document}

\title{The Absolute Age of M92}

\author{Jiaqi (Martin) Ying}
\affiliation{Department of Physics and Astronomy, Dartmouth College, 6127 Wilder Laboratory, Hanover, NH 03755, USA}

\author{Brian Chaboyer} 
\affiliation{Department of Physics and Astronomy, Dartmouth College, 6127 Wilder Laboratory, Hanover, NH 03755, USA}

\author{Emily M. Boudreaux}
\affiliation{Department of Physics and Astronomy, Dartmouth College, 6127 Wilder Laboratory, Hanover, NH 03755, USA}

\author{Catherine Slaughter} 
\affiliation{Department of Physics and Astronomy, Dartmouth College, 6127 Wilder Laboratory, Hanover, NH 03755, USA}
\affiliation{Leiden Observatory, Leiden University, NL-2300 RA Leiden, the Netherlands}

\author{Michael Boylan-Kolchin}
\affiliation{Department of Astronomy, The University of Texas at Austin, TX 78712, USA}

\author{Daniel Weisz}
\affiliation{Department of Astronomy, University of California Berkeley, CA 94720, USA}

\begin{abstract}
    The \textit{absolute age} of a simple stellar population is of fundamental interest for a wide range of applications but is difficult to measure in practice, as it requires an understanding of the uncertainties in a variety of stellar evolution processes as well as the uncertainty in the distance, reddening and composition. As a result, most studies focus only on the \textit{relative age} by assuming that stellar evolution calculations are accurate and using age determinations techniques that are relatively independent of distance and reddening. Here, we construct $20,000$ sets of theoretical isochrones through Monte Carlo simulation using the Dartmouth Stellar Evolution Program to measure the absolute age of the globular cluster M92. For each model, we vary a range of input physics used in the stellar evolution models, including opacities, nuclear reaction rates, diffusion coefficients, atmospheric boundary conditions, helium abundance, and treatment of convection. We also explore variations in the distance and reddening as well as its overall metallicity and $\alpha$ enhancement. We generate simulated Hess diagrams around the main-sequence turn-off region from each set of isochrones and use a Voronoi binning method to fit the diagrams to HST ACS data. We find the age of M92 to be $13.80 \pm 0.75$ Gyr. The $5.4\%$ error in the absolute age is dominated by the uncertainty in the distance to M92 ($\sim 80\%$ of the error budget); of the remaining parameters, only the total metallicity, $\alpha$ element abundance, and treatment of helium diffusion contribute significantly to the total error.
%
\end{abstract}

\section{Introduction} \label{Introduction}
Globular clusters (GCs) are stable, tightly bound clusters of stars. They are often modeled as simple stellar populations, as stars in a GC are believed to have the same origin and have similar composition and age. As a result, GCs are important observational basis for understanding composite stellar populations both inside and outside of Milky Way as they could be used as building blocks for the stellar population synthesis \citep{Bica86}. 

GCs are also the oldest objects in the Galaxy whose age may be accurately determined. Using JWST data, \citet{Mowla22} found globular clusters formed at $z > 9$, only $\sim 0.5$ Gyr after the big bang. Therefore, most GCs are relics of high-redshift star formation, and contain the fossil imprint of earliest phases of galaxy formation. Moreover, the bimodality of GCs (each galaxy generally has a metal-poor and a metal-rich sub-population) permit the investigation of two distinct phases of galaxy assembly (stellar halo and bulge) far beyond the Local Group \citep{Arnold11}. M92 is one of the oldest and most metal-poor galactic GCs \citep[e.g.][]{kraft_globular_2003}. The age of M92, therefore, provides a limit to the age of the universe \citep{chaboyer_age_1996} and insights into star formation in the early universe. 

Due to its richness, relative proximity and low reddening, M92 often serves as a benchmark for studies of low-metallicity stellar systems.  For example,  \citet{brown_quenching_2014} studied the stellar population of 6 ultra-faint dwarf galaxies (UFDs) using a combination of high-precision photometry data and pointed out that all 6 UFDs they studied showed that the stars of the smallest galaxies in the Universe was formed before reionization because the unusual similarity between their stellar population and that of the M92. As another example of the benchmark nature of M92, it was among the first objects observed with JWST as part of the early release science program 1334 \citep{Weisz23}. 

To first order, stars in a GC can be assumed to form at same time with the same composition; as a result, theoretical isochrone age fitting is the most widely-used method to determine the age of GCs. Theoretical isochrones can be generated by finding the common phase of stellar evolution shared by stellar evolution model with different mass \citep{dotter_dartmouth_2008}. A variety of methods have been applied to determine the best-fit isochrones for observational data and, therefore, constrain the age of GCs. 

Several examples exist for M92. \citet{salaris_homogeneous_2002} used the luminosity of the main sequence (MS) turn-off, combined with the color difference between the turn-off and the base of the giant branch, to find the age of M92 to be $12.3 \pm 0.9$ Gyr.  Using the luminosity of the main-sequence turn-off,  but a different distance estimate,   \citet{carretta_distances_2000} found the age of M92 to be $14.8 \pm 2.5$ Gyr. Utilizing the shape of the main sequence turn-off region as an age indicator, along with a new estimate of distance modulus (DM) and reddening, \citet{vandenberg_models_2002} estimated the age of M92 to be $13.5 \pm 1.0$ Gyr. \citet{cecco_absolute_2010} combined data from three different photometric systems --- Sloan Digital Sky Survey (SDSS), Johnson-Kron-Cousins, and Advanced Camera for Surveys (ACS) --- and, using the morphology, and number counts of stars in the main-sequence turn-off, red giant branch and horizontal branch, found the age of M92 to be $11 \pm 1.5$ Gyr. \citet{2009ApJ...694.1498M} derived ridge lines in Color Magnitude Diagram (CMD) from the ACS data to perform relative MS fitting between clusters and found the age of M92 to be $13.18 \pm 0.51$.  In general, the uncertainty in age determinations for GCs typically take into account the uncertainties in the observed properties of a globular cluster (distance, reddening, and composition), but do not include the uncertainty associated with the theoretical stellar models and isochrones which are used to determine the age.   

Modern stellar evolution codes can generate theoretical stellar models quickly for a  wide range of initial conditions. Theoretical isochrones are sensitive to the input parameters used to generate these stellar models.  Most previous studies using theoretical isochrones are limited in that they do not take into consideration the wide range of uncertainty in constructing stellar models. In this paper, we utilize a Monte Carlo (MC) approach to generate uncertainties in theoretical isochrones. We generate isochrones by varying various input physics in the stellar evolution models. For a given stellar model, we will prescribe its mass and heavy element composition and then vary the opacities, nuclear reaction rates, microscopic diffusion coefficients, atmospheric boundary conditions, helium abundance, and treatment of convection. All those parameters (shown in Table~\ref{tab1}) are varied during the MC simulation based upon their known uncertainties. The resultant isochrones provide a good estimation of the uncertainty associated with modern stellar evolution calculations. 

\begin{table*}[]
\caption{Monte Carlo Input parameters \label{tab1}}
\begin{tabular}{llll}
\hline
Variable                               & Distribution & Range        & Source                                               \\
\hline
{[}Fe/H{]}                    & Normal       & $-2.30 \pm 0.10$     &  \cite{Kraft2003}\\
 & & & \cite{Carretta2009} \\
 & & & \cite{Cohen2011} \\                                           \\
Primordial helium abundance   & Uniform      & $0.244 \sim 0.249$   & \citet{Aver2015}                                       \\
{[}$\alpha$/Fe{]}             & Normal       & $0.40 \pm 0.1$    & \cite{Roederer2011}                                               \\
Mixing length                 & Uniform      & $1.0 \sim 2.5$      & N/A                                         \\
Heavy element diffusion       & Uniform      & $0.5 \sim 1.3$      & \citet{Thoul1994}                                         \\
Helium diffusion              & Uniform      & $0.5 \sim 1.3$    & \citet{Thoul1994}                                           \\
Surface boundary condition 
 &   Trinary          &     1/3; 1/3; 1/3             & \citet{Eddington1926}\\
  &             &                     & \citet{KrishnaSwamy1966} \\
  &             &                     & \citet{Hauschildt1999} \\
Low temperature opacities     & Uniform      & $0.7 \sim 1.3$        & \citet{Ferguson2005}                                      \\
High temperature opacities    & Normal       & $1.0 \pm 0.03$           & \citet{Iglesias1996}     \\
Plasma neutrino loses         & Normal       & $1.0 \pm 0.05$     & \citet{Haft1994}     \\
Conductive opacities          & Normal       & $1.0 \pm 0.20$       & \citet{Hubbard1969}\\
 &             &                     & \citet{Canuto1970}   \\
Convective envelope overshoot & Uniform      & $0 \sim 0.2$      & N/A                                          \\
Convective core overshoot & Uniform & $0 \sim 0.2$ & N/A \\
$p + p \to H_2 + e + \nu$   
& Normal       & $\left(4.07 \pm 0.04 \right)\times 10^{-22}$ & \citet{Acharya16}\\
 & & & \citet{Marcucci13}\\
${ }^{3}He + { }^{3}He \to { }^{4}He + p + p$                 & Normal       & $5150 \pm 500$& \citet{adelberger_solar_2011}\\
${ }^{3}He + { }^{4}He \to { }^{2}H + \gamma$                  & Normal       & $0.54 \pm 0.03$&\citet{deBoer2014-rates}\\
${ }^{12}C + p \to { }^{13}N + \gamma$                & Normal       & $1.45 \pm 0.50$ & \citet{Xu2013}\\
${ }^{13}C + p \to { }^{14}N + \gamma$              & Normal       & $5.50 \pm 1.20$& \citet{Chakraborty2015}\\
${ }^{14}N + p \to { }^{15}O + \gamma$             & Normal       & $3.32 \pm 0.11$ & \citet{Marta11}\\
${ }^{16}N + p \to { }^{17}F + \gamma$               & Normal       & $9.40 \pm 0.80$ & \citet{adelberger_solar_2011}\\
\hline
\end{tabular}

\vspace{1ex}
{\footnotesize \textit{Note:} We adopt a $-3\,\sigma$ lower boundary for all physics-based parameters with normal distribution to prevent non-physical results (such as a negative nuclear reaction rate).  Units for nuclear reaction rates are keV-barns. Convective parameters are given in units of pressure scale heights. [Fe/H] and [$\alpha$/Fe] have their standard logarithmic definition with respect to the solar value.  The primordial helium abundance is given as a mass fraction. All other parameters are multiplicative about the standard value given in the source column.}
\vspace{0.5cm}
\end{table*}

A variety of methods have been used to compare theoretical isochrones to observational data in order to determine the age of a stellar population.  Typically, certain age-sensitive aspects of the observed color-magnitude diagram are compared to stellar models/isochrones. The main sequence turn-off region is particularly sensitive to age and is therefore often used to determine the ages of GCs \citep[e.g.][]{vandenberg_models_2002}. However, the morphology of the horizontal branch has also been used in measuring GC ages \citep[e.g.][]{salaris_homogeneous_2002, cecco_absolute_2010}.  \citet{vandenberg2016} used both the main sequence turn-off region and the horizontal branch to determine the age a few globular clusters, including M92.  These previous studies have focused on comparing the shape/morphology of the observational data to theoretical isochrones in order to determine their age.

In this paper, we present a new isochrone age fitting method which uses Voronoi binning and fits the number \textit{density} of stars in the main sequence turn-off region to determine ages. By utilizing the density of stars in the color-magnitude diagram (usually referred to as a Hess diagram) to determine the age of a cluster, we utilize all of the observational information to constrain the age.  This may lead to smaller uncertainties compared to previous work.  This paper is structured as follows. In \S \ref{Observational Data}, we introduce the observational data;  \S \ref{Isochrone Construction} covers the process of isochrone construction; \S \ref{Isochrone Fitting} presents the details our isochrone age fitting method and our best age measurement; and \S \ref{sec:discussion} includes a discussion of the sources of error and covariance.

\section{Observational Data} \label{Observational Data}

\subsection{Calibration Stars} 
There are two single, metal-poor main sequence stars that have accurate HST ACS photometry (in the same filters as the M92 ACS data) and virtually the identical composition as M92: HIP 46120 and HIP 106924.  The HST photometry is presented in \citet{Chaboyer2017} and the high resolution spectroscopic abundances in \citet{omalley2017}. These two stars have accurate $Gaia$ EDR3 parallaxes of $	14.776\pm 0.014\,$mas and $15.019\pm 0.012\,$mas respectively \citep{EDR3, EDR3pi}. 

The EDR3 parallaxes are known to suffer from systematic zero-point errors, and a calibration of this error has been given by   \citet{EDR3bias}.  This correction is $-0.012\,$mas for HIP 46120 and $-0.020\,$mas for HIP 106924.  However, the zero-point correction is not that well calibrated for bright stars like these two stars \citep{EDR3bias} and there is evidence that the zero-point correction may be an over-correction for bright stars \citep{riess2021,zinn2021} so we elected to add in half the zero-point correction to the quoted EDR3 parallax.   The uncertainty in the parallaxes was taken to be the value of the zero-point correction added in quadrature with the uncertainty in parallax given in EDR3. 

Combining the parallaxes with the HST ACS photometry, we measure $M_{F606W} = 5.7867\pm 0.0026\,$mag for HIP 46120 and $M_{F606W} = 6.0406\pm 0.0037\,$mag for HIP 106294.  These stars have zero reddening \citep{omalley2017} and observed colors of (F606W-F814W)$ = 0.566\pm 0.002$ and  (F606W-F814W)$ =0.601\pm 0.005$. These accurate colors and absolute magnitudes will be used to test the isochrones in \S \ref{Calibration Star test}.

\subsection{M92} 
To estimate the age of M92, we use calibrated data for M92 from the Hubble Space Telescope (HST) Advanced Camera for Surveys (ACS) globular cluster survey treasury program \citep{sarajedini_acs_2007,  anderson_acs_2008}. The survey obtained photometry with S/N $>10$ for main sequence stars with masses $> 0.2\,M_{\odot}$ using the ACS Wide Field Channel. Artificial star tests provide an accurate estimate of the photometric uncertainties and completeness as a function of magnitude and cluster position \citep{anderson_acs_2008}. Since this paper focus on determining the age of M92, we use a subset of stars around the main sequence turn-off to fit isochrones whose position is most sensitive to variations in age, and relatively insensitive to the present day mass function.  These stars have a $15.925 < \mathrm{F606W} < \mathrm{19.925}$, which is $\pm 2\,$magnitudes of the point on the subgiant branch which is $0.05\,$mag redder than the main sequence turn-off (MSTO). Additionally, we remove blue straggler stars and outliers by selecting stars that are within $0.08\,$mag in F606W of the median ridgeline in an magnitude-magnitude diagram of F814W and F606W. With these cuts, our observational sample contains  $18,077$ stars.  

We note that previous studies (e.g., \citealt{miloneHubbleSpaceTelescope2017, meszarosExploringAnticorrelationsLight2015}) demonstrate that M92, like other old globular clusters, hosts multiple stellar populations. These multiple stellar populations typically have somewhat different lighter element abundances, the origin of which is not currently known.  These multiple populations are observed in UV filters such as F275W and F336W from the HST. However, these populations are indistinct from each other in the F606W and F814W data used in this paper. As a result, these multiple populations will not be considered in this study.

\section{Isochrone Construction and Testing} \label{Isochrone Construction}

We use the Dartmouth Stellar Evolution Program (DSEP) \citep{dotter_dartmouth_2008} to generate stellar models and isochrones and generally use literature estimates when adopting uncertainties for each parameter (see Table~\ref{tab1}). One area where we consider a wider range of uncertainties is in the treatment of convection: even though nearly all models use a solar-calibrated mixing length, a variety of studies have demonstrated that this may not be the most appropriate value for other stars. \citet{joyceNotAllStars2018} studied metal-poor stars including M92 and discovered that solar-calibrated value of the mixing length parameter $\alpha_{\textup{MLT}}$ was ineffective at reproducing their observed properties. As a result, we adopt a wider input range for the mixing length parameter $\alpha_{\textup{MLT}}$ to cover the range of empirical calibrated mixing length parameter $\alpha_{\textup{MLT}}$. Another source of uncertainty associated with the treatment of convection in stellar models is the amount of convective overshoot, which may occur at the formal edge of a convection zone (which is defined by the buoyancy force being zero).  Various studies have calibrated the amount of convective overshoot by comparing to observations (e.g., \citealt{demarqueY2IsochronesImproved2004, claretNewGridsStellar2004, pietrinferniLargeStellarEvolution2004, mowlaviStellarMassAge2012}). In general, these studies have found a fairly small value of 0.0 to 0.2 pressure scale heights; we therefore adopt this range for convective overshoot in our analysis.

\begin{figure}
    \centering
    \includegraphics[width=0.45\textwidth]{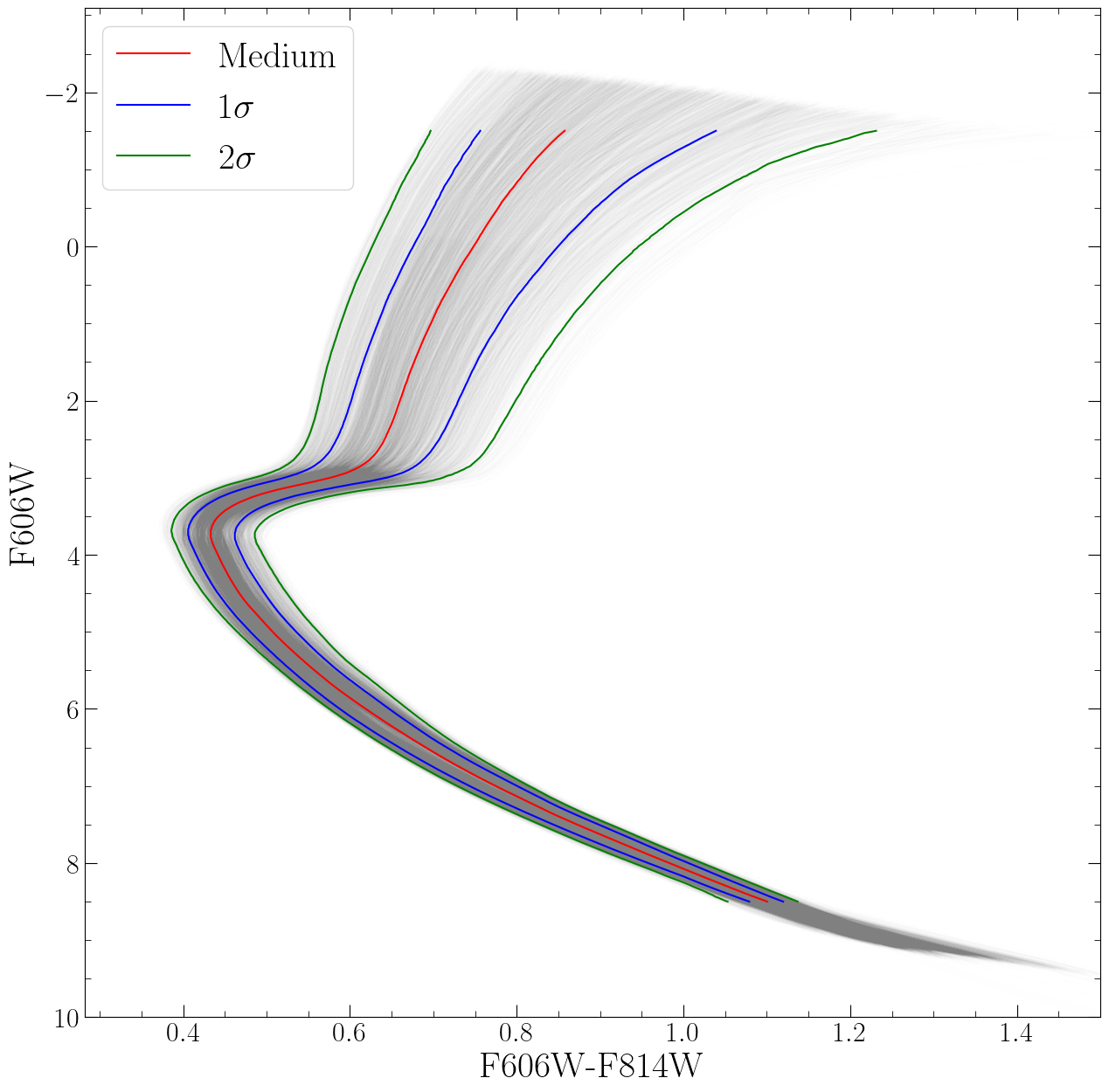}
    \caption{An illustration of the $20,000$ 13 Gyr MC-isochrones (Grey) generated for this project. Red curve is the interpolated isochrone which is selected by taking the median F606W-F814W value of all isochrones at a given F606W value. The corresponding $68 \%$ confidence intervals and $95 \%$ confidence intervals are shown in blue and green curves, respectively.}
    \label{fig:13_gyr_iso}
\end{figure}

We generate $20,000$ sets of input parameters by doing Monte Carlo simulations on parameters shown in Table \ref{tab1} from their associated probability distribution functions. Each set of input parameters is used to evolve 21 stellar models with mass from $0.65\,M_{\odot}$ to $1.5\,M_{\odot}$ with an increment of $0.05 \,M_{\odot}$ and 12 lower-mass stellar models with mass from $0.3\, M_{\odot}$ to $0.63 \,M_{\odot}$ with an increment of $0.03 \,M_{\odot}$. The lower-mass models use the FreeEOS-2.2.1 \citep{irwinFreeEOSEquationState2012}, while the higher-mass models use an analytical equation of state which include the Debyre-Huckel correction \citep{chaboyerOPALEquationState1995}.  These stellar models are used to generate 41 theoretical isochrones from $8\,$Gyr to $16\,$Gyr with an increment of $200\,$Myr. Those $41$ isochrones of different ages corresponding to the same set of MC input parameters, are considered a single MC isochrone set. Each isochrone is  constructed with a dense grid of 400 equal evolutionary points in order to ensure that the output isochrones have a high density of points to avoid any interpolation errors when constructing simulated color-magnitude diagrams (sCMDs)\footnote{The Monte Carlo isochrones created for this project are available at \url{https://doi.org/10.5281/zenodo.7758605}. The file is stored in HDF5 format, and a sample python program is provided which gives details on  on extracting isochrones from the HDF5 file.}. In summary, we generated $20,000$ isochrone sets.  Each isochrone set consists of  $41$ isochrones of different ages, for a total of $20,000 \times 41 = 820,000$ individual isochrones. Figure \ref{fig:13_gyr_iso} shows the distribution of all $20,000$ 13-Gyr isochrones generated for this project. The extensive range covered by the single age isochrones in the color-magnitude plane affirms our hypothesis that varying the MC input parameters can significantly influence the resulting isochrones. Hence, it is imperative to take into account the uncertainty in these parameters when provide an accurate determination of the age of M92.

\subsection{Testing the Isochrones}\label{Calibration Star test}
As HIP 46120 and HIP 106924 have known absolute magnitude and colors, and nearly identical compositions to M92, these stars provide an empirical baseline for an isochrone goodness-of-fit metric. Specifically, we preform a $\chi^{2}$ goodness-of-fit test between the two calibrating stars and each age of each MC isochrone set. 
The lowest $\chi^{2}$ value for a given MC isochrone set is then used to  compute a weighting function for that entire set of MC isochrones (i.e., for a given MC isochrone set, we assume the best-fitting isochrone gives an indication of the age of the calibrating star, which may be different from the age of M92). Essentially, how well any given MC isochrone set fits to the observed calibration data will determine the weight that a MC isochrone from that set is given when fitting to M92. We use the inverse of the probability that a given MC isochrone is inconsistent with the calibration data as the weighting function.

We define the $\chi^{2}$ metric for of each isochrone as the quadrature sum of the differences between that isochrone's and calibrating star's F606W magnitude and F606W-F814W color. In order to account for uncertainty in the calibrating stars photometry the differences used are normalized by the magnitude and color uncertainties. $\chi^{2}$ is found for each age in each of the 20,000 sets of isochrones. The age with the minimum $\chi^{2}$ is then selected for each set of isochrones. From these minimized $\chi^{2}$ values, we directly compute the weighting function. 

\begin{figure}
    \centering
    \includegraphics[width=0.45\textwidth]{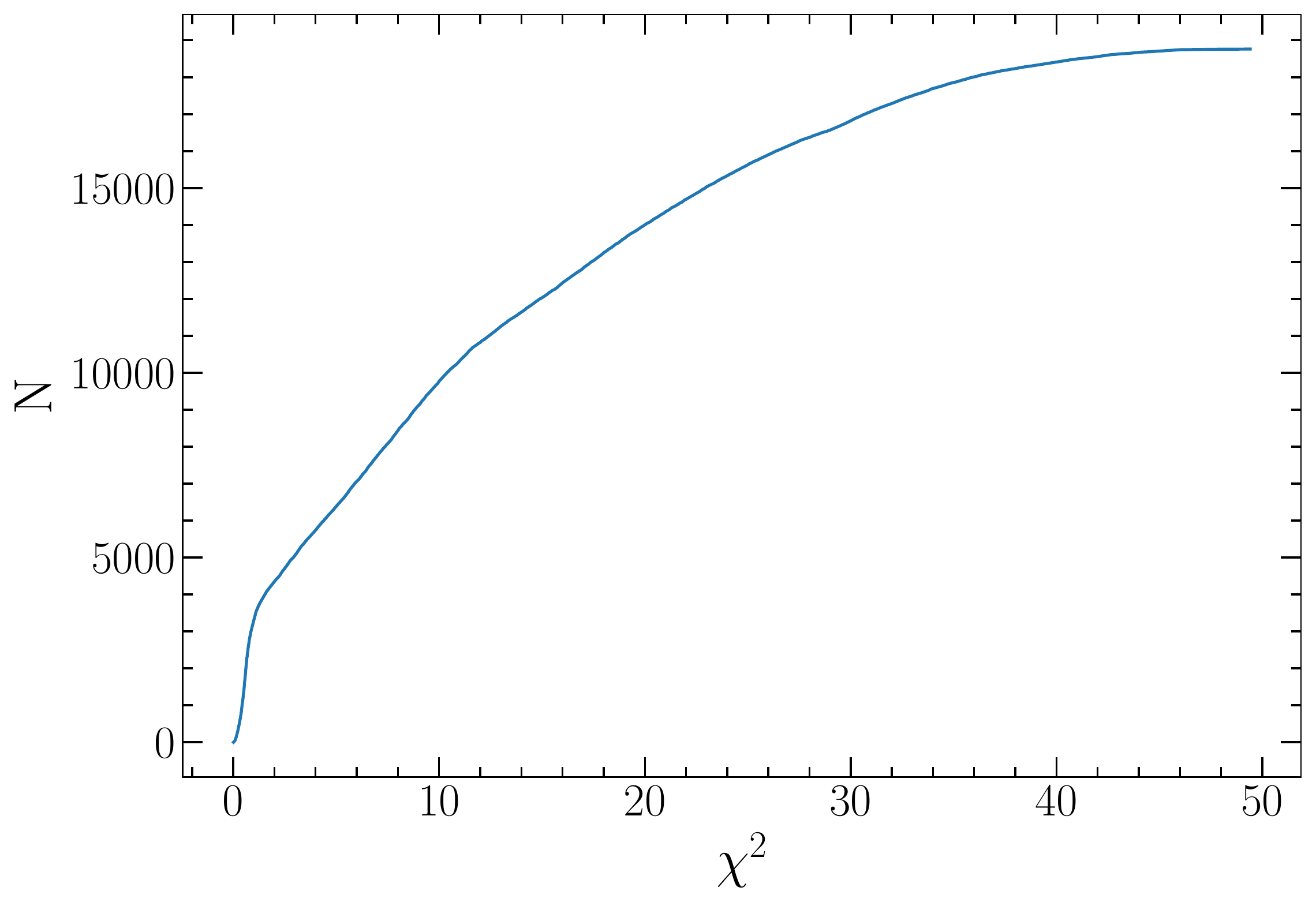}
    \caption{The cumulative distribution of the minimum  $\chi^{2}$ values for all MC isochrones. }
    \label{fig:chipdist}
\end{figure}

\begin{figure}
    \centering
    \includegraphics[width=0.45\textwidth]{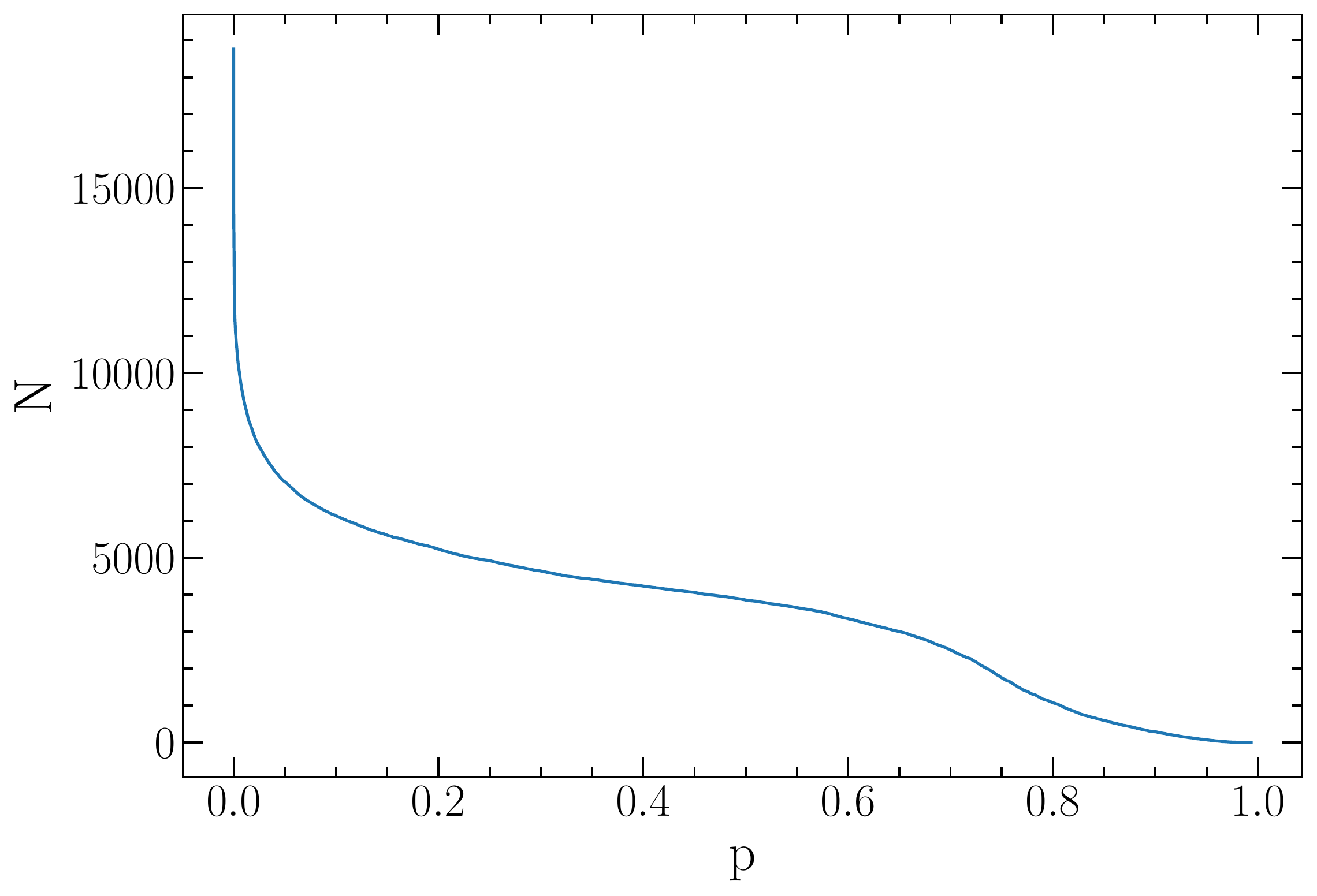}
    \caption{The distribution of the weight $p$ for $20,000$ sets of MC isochrones.}
    \label{fig:cdfdist}
\end{figure}

The two calibrating stars provide 2 degrees of freedom, $n$, for the $\chi^{2}$ distribution; in the case where $n=2$, the cumulative distribution function (CDF) for a $\chi^{2}$ distribution is given by 
\begin{equation}
\label{eqn:cdfchi2n2}
    \text{CDF} = 1 - e^{-x/2}\,.
\end{equation}
The weighting function used, $p$, for any given isochrone is then 
\begin{equation}\label{eqn:weightFuncCalibrate}
    p = 1-\text{CDF}\,.
\end{equation}
Figure \ref{fig:chipdist} shows the cumulative distribution of minimized $\chi^{2}$ values, and Figure~\ref{fig:cdfdist} shows  the cumulative distribution of the weight $p$ for $20,000$ sets of MC isochrones.  Approximately  $20\%$ of the isochrone sets  have $\chi^2 < 2$ and so provide a good fit to the calibration stars.  Only about $35\%$ of the isochrone sets  have a CDF $< 0.95$, which corresponds to a weighting function $> 0.05\%$.

\subsection{Simulated Color-Magnitude Diagram} \label{Simulated Color-Magnitude Diagram}

Each MC set of theoretical isochrones are used to create a set of simulated color-magnitude diagrams (sCMDs)  of M92 which will be used to compare with the observational CMD of  \cite{sarajedini_acs_2007}. A sCMD is constructed by randomly creating a four million point sample for each isochrone in the following steps:
\begin{enumerate}
    \item A random distance from the center of the cluster is selected from the observed distribution \citep{sarajedini_acs_2007}.
    \item A random mass is selected using the present day mass function determined by \citet{paust_acs_2010}, who found a power law mass function with a slope of $-1.02$ using the same ACS M92 data.  The magnitudes (F606W and F814W) of this simulated star are then determined from the isochrone. 
    \item The simulated star is randomly assigned to be a member of a binary system, using the observed binary mass fraction of $0.02$ \citep{milone_acs_2012}. 
    \item If a star is a member of a binary system, then a secondary star is created, assuming a  flat secondary mass distribution with mass ratio  $q=0.5$ to $1.0$.  The magnitudes of this secondary star is determined from the isochrone and added to the magnitude of the primary to arrive at the magnitude of the binary star system; which is considered to be a single star in the photometry.  
    \item It is determined if the star would be recovered in the  photometric reduction, using the photometric completeness function from \citep{anderson_acs_2008} for the M92 ACS data.  This photometric completeness function depends on the magnitudes of the star, and its distance from the center of the cluster.
    \item If the star is found to be observable in the previous step, then photometric errors will be randomly selected from their observed distribution (which is a function of magnitude and distance from the cluster center).  The observed distribution of photometric errors is determined from the artificial star tests of \citet{anderson_acs_2008}.  
    \item The photometric error in F606W and F814W are added to the magnitude of the simulated star, and these magnitudes are used in creating the sCMD.  
\end{enumerate}

Once the sCMD is created, the same colour and magnitude filters which were applied to the observed M92 data are applied to the sCMD.  After this filtering, the sCMD  consists of about two million simulated data points for each theoretical isochrone.

We note that after we completed our age determination for M92, we discovered that \cite{2020MNRAS.494.4226E} found a strong correlation between the distance from the center the GC and the present-day mass function (PDMF) slope. \cite{2020MNRAS.494.4226E} found a PDMF slopes in the inner region of M92 which ranged from $-1$ to $+2$, while our sCMD assumed $-1.02$  \citep{paust_acs_2010}.  Since we are only fitting stars which are around the main-sequence turn-off, which all have similar masses, the exact PDMF slope should have little impact on our results.  To test this, we created sCMDs with PDMF slopes ranging from $-2.02$ to $1.02$ and found the change in PDMF slope indeed had a negligible impact on the estimated age for M92. 

\section{Isochrone Fitting} \label{Isochrone Fitting}

To estimate the age of M92, each sCMD is compared to the observational CMD and the fit probability is calculated. In order to compare CMDs, we divide the 2D CMD into multiple subsections and estimate the goodness of fit using a $\chi^2$ method:
\begin{equation}
\chi^2 = \sum_{i} \frac{\left( O_i - E_i \right)^2}{E_i} \label{eq1},
\end{equation}
where $E_i$ is the number of data points in a subset of the CMD in the observational data and $O_i$ is the number of sampled data points in the same subset from the sCMD. Since the number of stars in the sCMD is $\sim 100\,$times larger than the number of stars in the observed CMD, the uncertainty in the number counts for simulated stars is negligible in comparison  to the uncertainty in the number counts for the observed stars.  The age determination is done in a series of steps, as discussed below. 


\subsection{Voronoi Binning} \label{Voronoi Binning}
To compare the sCMD with the observational CMD, a method to partition the 2D CMD was required. The most intuitive method will be dividing the CMD using a uniform grid. However, the distribution of stars in the CMD is highly biased. As the result, if the bins of a 2D CMD were defined by evenly spaced grids, there would be a wide distributions of expected data points in each bin, with some bins being empty (either in the real, or simulated data).  Therefore equation (\ref{eq1}) could not be used for that bin. A better approach of a non-uniform partition of the 2D CMD result in roughly equal number of points per bin is required.  

To achieve this requirement, we use the adaptive Voronoi binning method of \citet{cappellari_adaptive_2003}. The algorithm sets up initial bins based on a Voronoi Tessellation formed by the simulated CMD. It iteratively combines bins nearby and thus raises the signal to noise ratio until it reaches the target. It satisfies  three requirements:
\begin{enumerate}
\item Topological requirement: there will be no data points which are not in a bin, and no bins overlap, 
\item Morphological requirement: the bin shape will be as ``compact'' (or ``round'') as possible so that two pixels from two corners across the CMD will not be put into one bin, 
\item Uniformity requirement: the resulted bins will have similar number of data points as targeted. Therefore, all bins can be considered as having equal statistical significance.
\end{enumerate}

This Voronoi binning is extremely computational demanding, and to save CPU time, $180,000$ points from each sCMD are randomly selected to generate Voronoi bins. Each set of Voronoi bins contain $800$ bins with average of $225$ points in each bin. Because the Voronoi binning method of \citet{cappellari_adaptive_2003} was designed to deal with images and required the pixel-size to be the same on both axis, we rescaled the magnitude of F606W-F814W  before doing the Voronoi binning. Because the linear transformation does not change the topology of the data, the original data can be easily recovered for further analysis. Different combinations of number of data points used and number of Voronoi bins were tested and the combination being used in this paper shows a balance in computational time and accuracy. After the Voronoi bin are determined, the entire sCMD will be binned and contribute to the expectation in equation (\ref{eq1}).





\begin{figure}[h]
\centering
\includegraphics[width=0.45\textwidth]{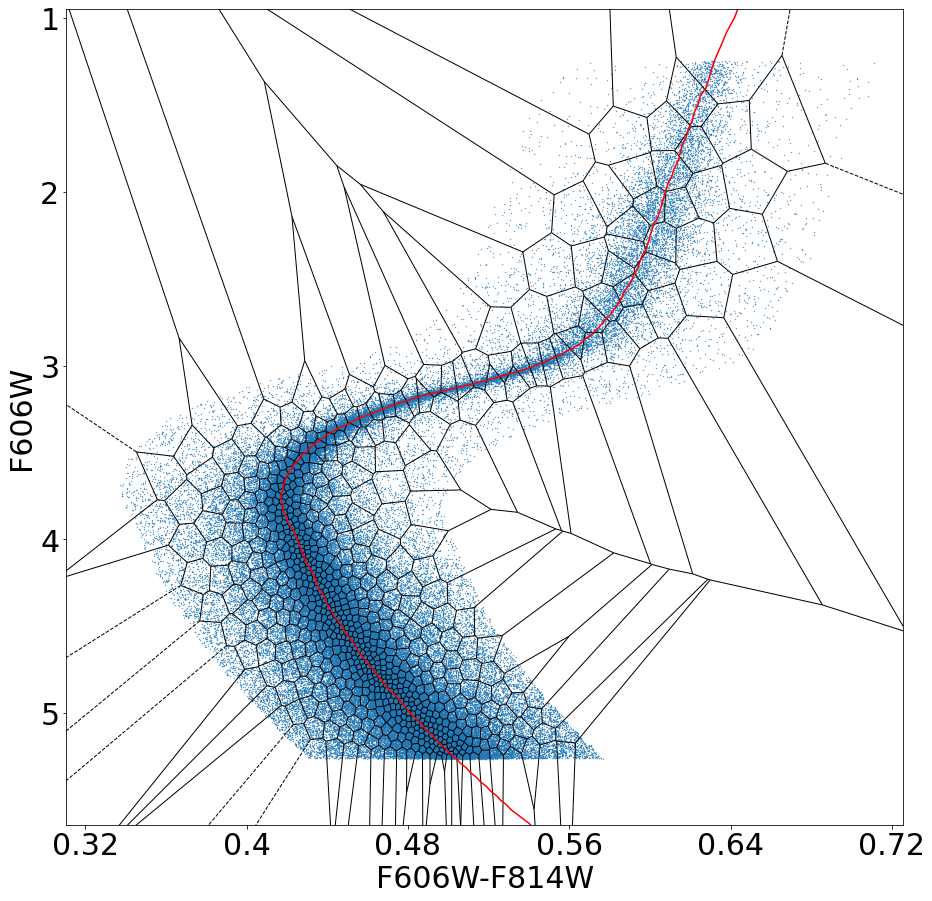}
\caption{A example of the result of the adaptive Voronoi binning method \citep{cappellari_adaptive_2003}. The CMD is divided into $800$ different subsets. Blue dots are the $180,000$ data points used to generate the Voronoi tessellation. Red curve is the isochrones used to simulate the CMD. } 
\label{fig1}
\end{figure}

Figure \ref{fig1} shows an example of the Voronoi binning \citep{cappellari_adaptive_2003}. The CMD is divided into $800$ different subsets with different sizes. Most bins locate near the isochrone (red line) and few bins locate outside where density of simulated data points (blue dots) is low.

For each set of Voronoi bins, the observational CMD is shifted by a range of distance modulus and reddening, with the ranges chosen to encompass the observed uncertainties in these quantities.  The distance to M92 is estimated using main sequencing fitting using the two calibration stars which good HST photometry and parallaxes from Gaia EDR3 \citep{collaborationGaiaEarlyData2021}.  Assuming a reddening of $E_{F606W-F814W} = 0.02$, the distance modulus of $(m-M)_{F606W}= 14.80 \pm 0.02$ is found. Assuming a reddening of $E_{F606W-F814W} = 0.01$, the main sequence fitting yields a  distance modulus of $(m-M)_{F606W}= 14.75 \pm 0.02$.  The evidence favours the higher reddening value.  An independent distance estimate to M92 by \citet{vandenberg_models_2002} is $(m-M)_{V}= 14.62$ and $E_{B-V} = 0.023$ based upon fitting ground based data to their isochrones (which assumes no uncertainty in their isochrones).  \citet{cecco_absolute_2010} found a distance modulus of $(m-M)_{V}= 14.82$ and $E_{B-V} = 0.025$ from fitting an different set of isochrones to a different ground based dataset.   \citet{baumgardtAccurateDistancesGalactic2021} used a variety of methods (including EDR3 parallaxes of cluster stars, and main sequence fitting) to estimate  distance to M92 to be $8.48 \pm 0.17\,$kpc which is $(m-M)_{o}= 14.64 \pm 0.04$. Assuming a reddening of $E_{B-V} = 0.01$, this corresponds to $(m-M)_{F606W}= 14.67$.  Based upon the above, the distance modulus used in this paper range from 14.62 to 14.82 with an increment of 0.01, and reddening range from 0.0 to 0.05 with an increment of 0.01. For each combination of distance modulus and reddening, a $\chi^2$ value was calculated using equation (\ref{eq1}) for each of the 41 ages in each MC isochrone set.  The minimum $\chi^2$ was then selected as the best estimate of the age, distance modulus and reddening for that particular MC isochrone.  



\subsection{Empirical $\chi^2$ Distribution} \label{Empirical chi2 Distribution}

\begin{figure}[t]
    \centering
    \includegraphics[width=0.45\textwidth]{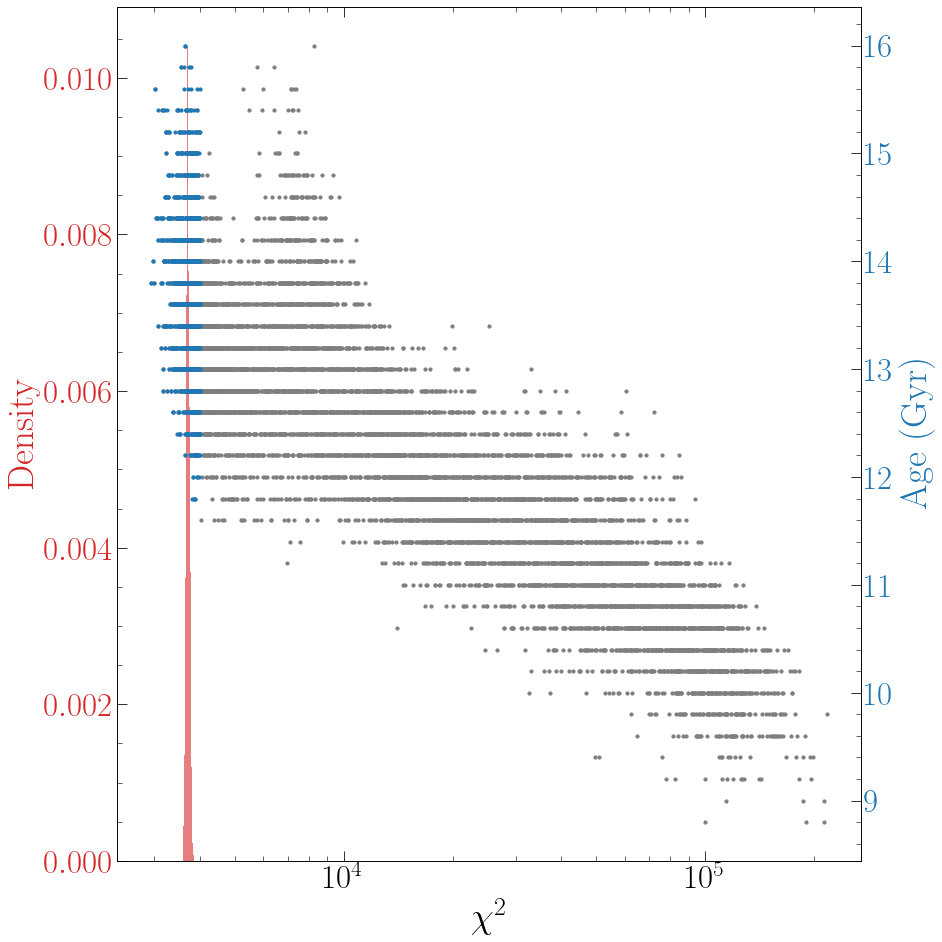}
    \caption{The distribution of $\chi^2$ values found using eqn.~(\ref{eq1}) when comparing the sCMDs to the observed data.  The empirical $\chi^2$ distribution is shown in red, and the $\chi^2$ values found when fitting the M92 data to the MC isochrones are shown as a function of age (scale shown on the right vertical axis).  The $1,100$ isochrones which are  within $3\,\sigma$ of the mean of the empirical distribution are shown as blue dots. The remaining isochrones have $\chi^2> 4100$ and are shown as grey dots. Note that the  x-axis is on a logarithmic scale.}
    \label{fig:ages}
\end{figure}

\begin{figure}[t]
    \centering
    \includegraphics[width=0.45\textwidth]{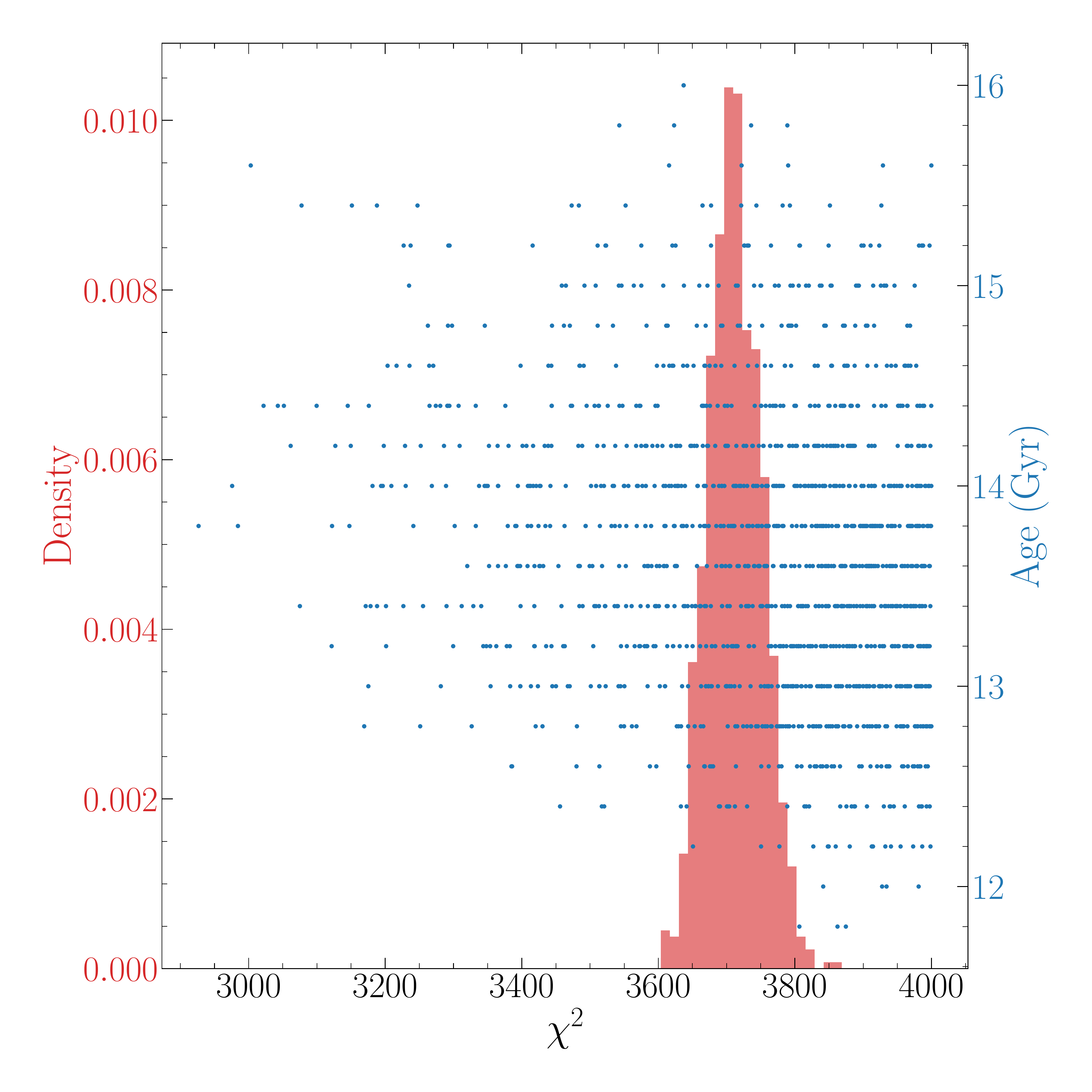}
    \caption{The empirical $\chi^2$ distribution is shown in red, while the $\chi^2$ values determining from fitting the MC isochrones to M92 are shown in blue (as a function of age). Only the $1,100$ isochrones which are  within $3\,\sigma$ of the  the mean of the empirical distribution are shown in this figure.  The x-axis is in linear scale.}
    \label{fig:ages_2}
\end{figure}

\begin{figure*}[htp]
    \centering
    \includegraphics[width=0.8\textwidth]{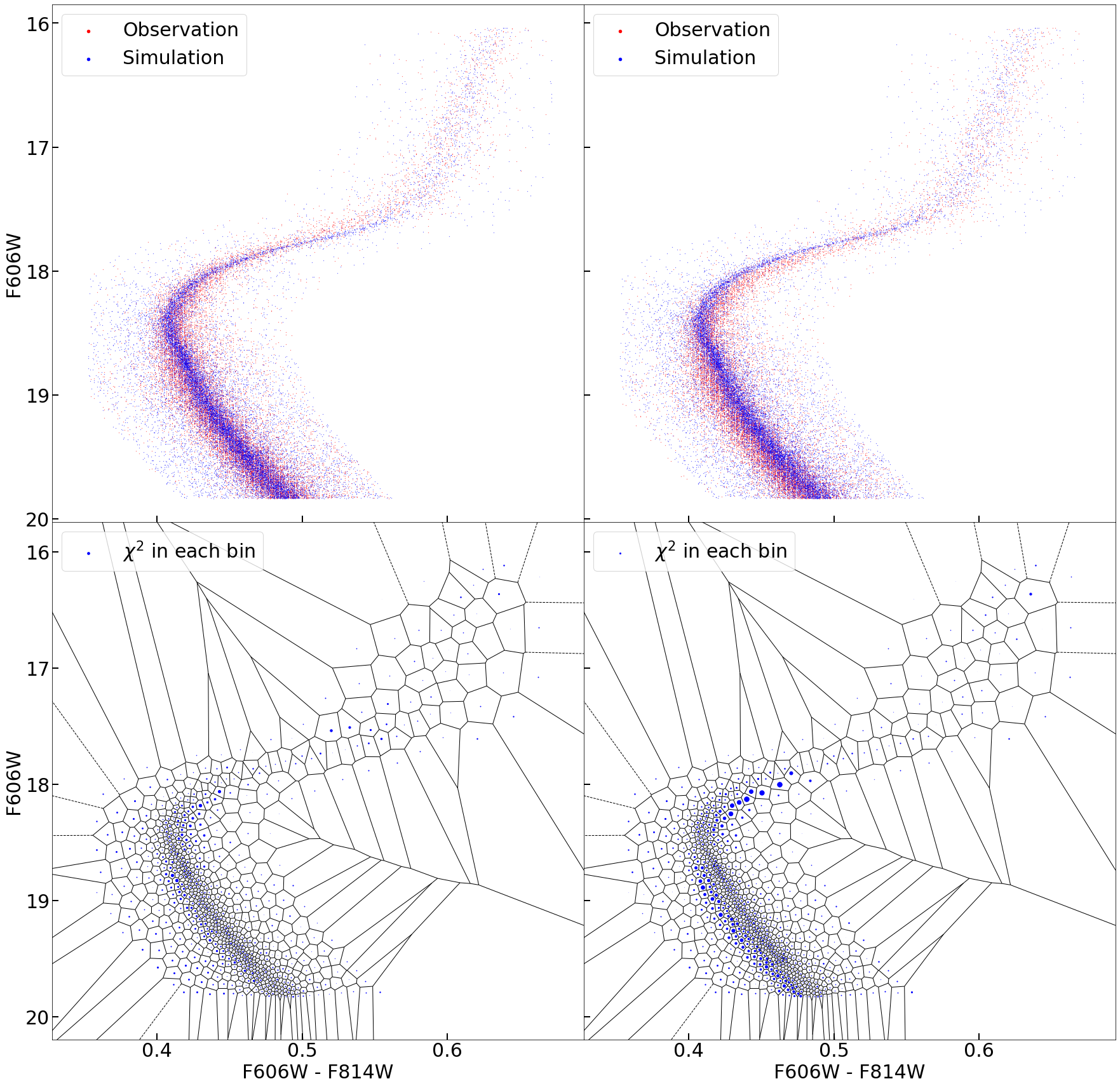}
    \caption{Top left (a): An example of ``high probability fit" for M92. The $\chi^2$ value is within $1\,\sigma$ of the mean of the empirical distribution. Blue dots are parts of the sCMD and red dots are observed data for M92. 
    Top Right (b): An example of ``low probability fit" for M92. The $\chi^2$ value is almost $3\,\sigma$ greater than  the mean of the empirical distribution. Blue dots are parts of the sCMD and red dots are observed data for M92. 
    Bottom left (c): An example of ``high probability fit" for M92. The size of the blue dots indicates the $\chi^2$ value for the Voronoi bin they located at.
    Bottom Right (d): An example of ``low probability fit" for M92. The size of the blue dots indicates the $\chi^2$ value for the Voronoi bin they located at.}
    \label{fig:test}
\end{figure*}

\citet{linResearchCommentaryToo2013} 
demonstrated that with large data sets, using the p-value-based hypothesis testing method no longer provides scientifically reliable results. Therefore, with the M92 data, it is  inappropriate to estimate the goodness of fit using the standard $\chi^2$ fit probability function.  To interpret $\chi^2$ values calculated in section \ref{Voronoi Binning}, a statistical method to determine the empirical $\chi^2$ distribution is required.  To do so, we re-sample the observational data using the photometric error and completeness from the artificial star test \citep{anderson_acs_2008}. From the observed data, $10,000$ CMDs each with about two million data points are generated. Using the same method described in section \ref{Voronoi Binning}, for each re-sampled CMD, a set of Voronoi bins is determined and a $\chi^2$ value is calculated using equation (\ref{eq1}). As a result, an empirical $\chi^2$ distribution is determined, and is used to compare with theoretical values.


Figure \ref{fig:ages} shows the empirical $\chi^2$ distribution and the $\chi^2$ values determined when comparing the MC isochrones to the observations. From the $20,000$ sets of theoretical isochrones created in this study, $1,100$ isochrones are within $3\,\sigma$ of the mean of the empirical distribution. The other $18,900$ MC isochrones yielded a very poor fit to the observed data. Figure \ref{fig:ages_2} is a zoomed-in version of Figure \ref{fig:ages} and shows $1,100$ isochrones are within $3\,\sigma$ of the mean of the empirical distribution. Most of these $1,100$ isochrones also provided a good fit to the calibration star data.  However, $66$ of the isochrones provided relatively poor fits to the calibration stars, leading to a calibration star weighting value of less than 0.10.   The shape of the empirical $\chi^2$ distribution can be fit with a normal distribution $\chi^2 = 3712 \pm 39$ and spans a very narrow region in $\chi^2$. Figure \ref{fig:ages} shows that the mean of the empirical $\chi^2$ distribution can be several orders of magnitude smaller than the $\chi^2$ of isochrones which poorly fits the data, which shows the sensitivity of isochrones to MC parameters and the selectivity of our age determination technique. Figure \ref{fig:test}(a) and Fig. \ref{fig:test}(b) shows examples of the sCMDs which are generated from those $1,100$ theoretical isochrones. Figure \ref{fig:test}(a) has a $\chi^2$ value which is within $1\,\sigma$ of the mean of the empirical distribution and is considered as a ``high probability fit" for M92, with an  higher weight in the final age estimation. Figure  \ref{fig:test}(b) has a  $\chi^2$ value which is almost $3\,\sigma$ greater than  the mean of the empirical distribution and is considered a ``low probability fit" for M92. Although it is taken into consideration in the age estimation, it had a much lower weight.  

Figure \ref{fig:test}(c) and Fig. \ref{fig:test}(d) are the corresponding $\chi^2$ values for each of the Voronoi bins of the two sCMDs shown in Fig. \ref{fig:test}(a) and Fig. \ref{fig:test}(b), respectively. Figure \ref{fig:test}(c) and Fig.\ref{fig:test}(d) shows a difference mostly in main sequence and MSTO which favors the sCMD shown in Fig. \ref{fig:test}(a). 

\subsection{Age Estimation} \label{Age Estimation}

\begin{figure}[htp]
    \centering
    \includegraphics[width=\columnwidth]{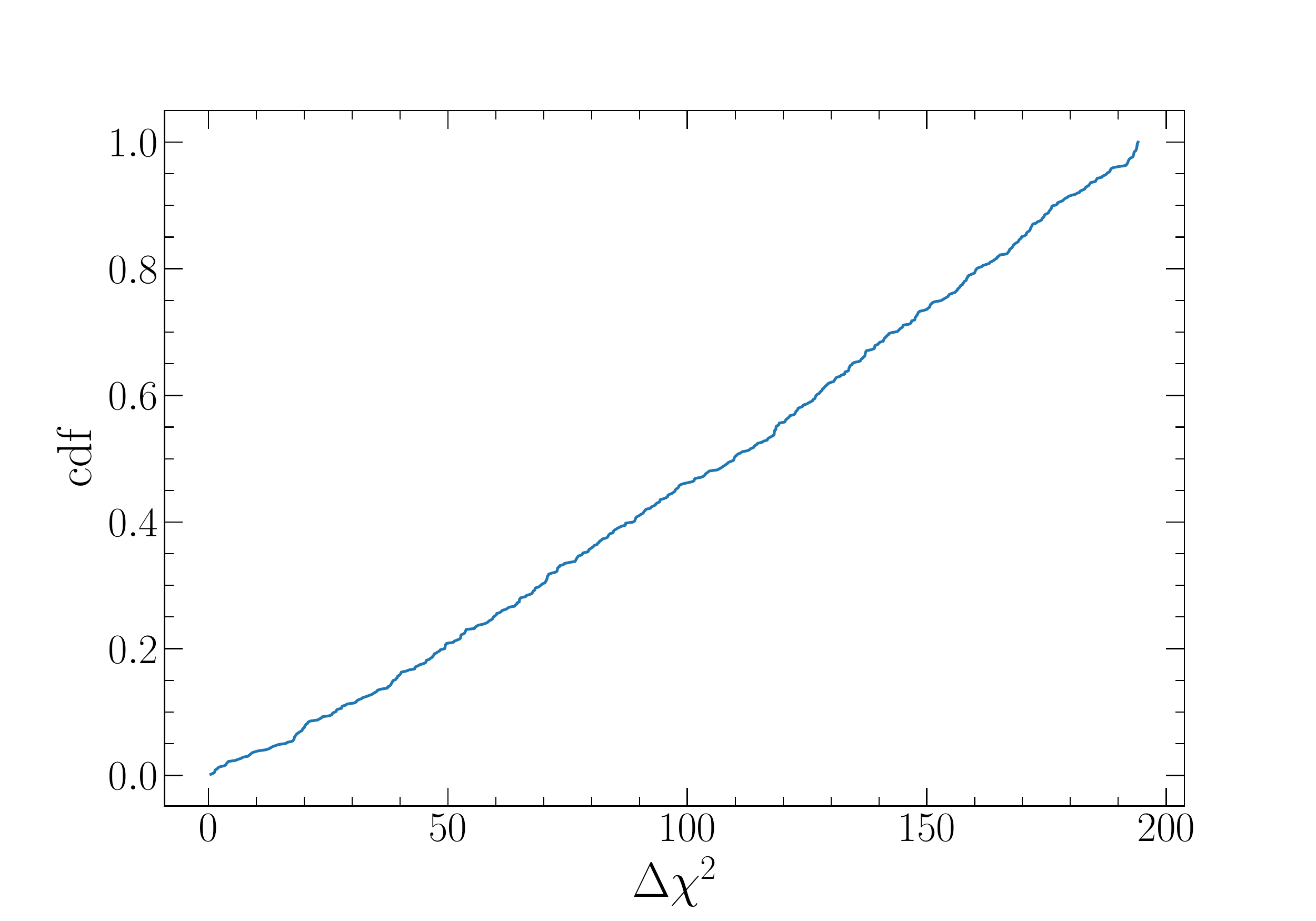}
    \caption{The cumulative distribution function of the $\chi^2$ values which are higher than the mean of the empirical $\chi^2$ distribution while are within $3\,\sigma$ of the  the mean of the empirical distribution.}
    \label{fig:delta_chi2}
\end{figure}

 Figure \ref{fig:ages} shows a clear bias towards $\chi^2$ values higher than the mean of the empirical $\chi^2$ distribution. As a result, the $\chi^2$ values smaller than the mean of the empirical $\chi^2$ distribution was considered to have a ``fit probability" of 1 while the ``fit probability" of $\chi^2$ values higher than the mean of the empirical $\chi^2$ distribution was defined by the empirical $\chi^2$ distribution where the cumulative distribution function is shown in Fig. \ref{fig:delta_chi2}. 

\begin{figure}[htp]
    \centering
    \includegraphics[width=\columnwidth]{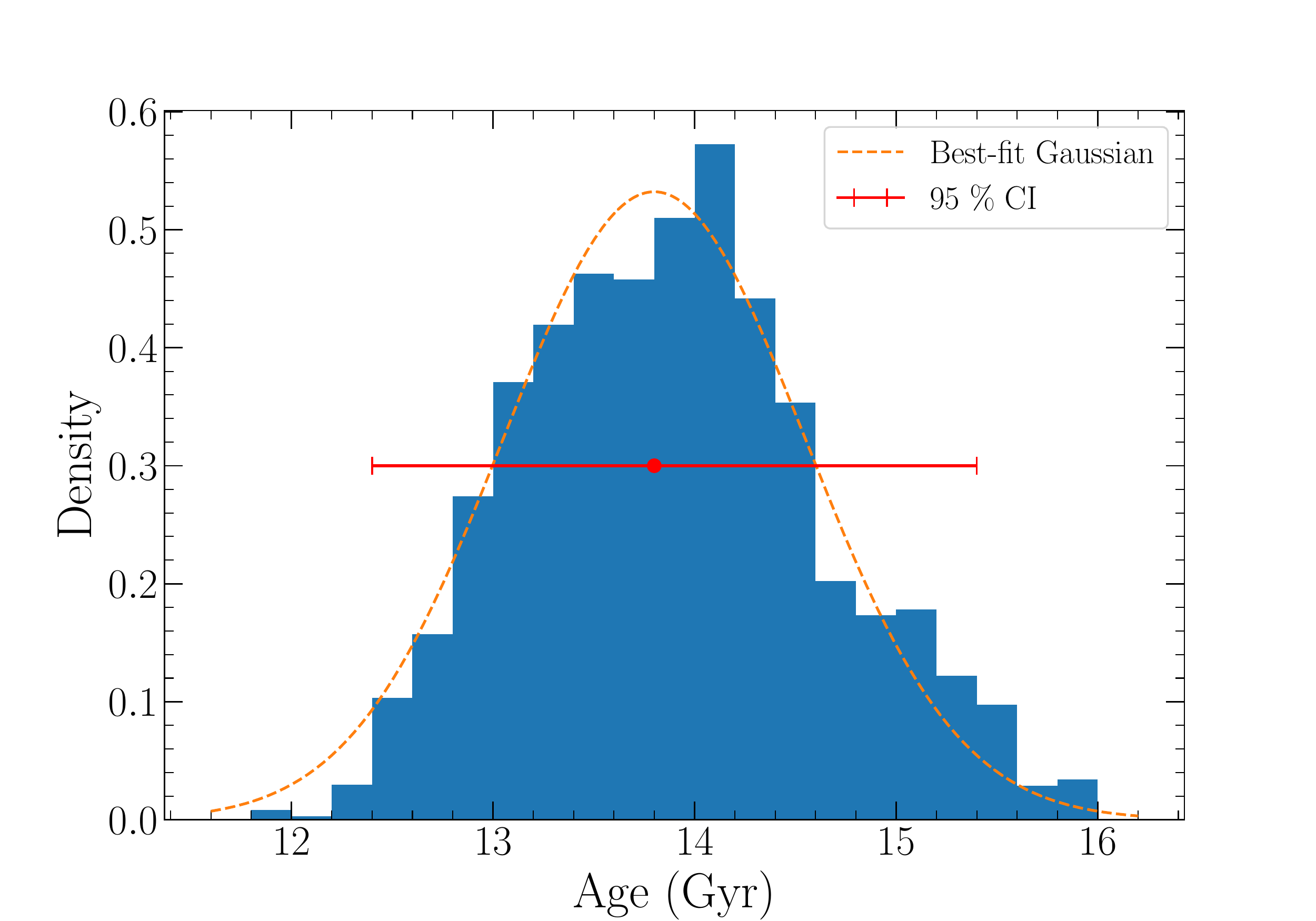}
    \caption{The weighted distribution of ages corresponding to best-fit isochrones (blue histogram). The orange curve represents the best-fit Gaussian model, $\mathcal{N}(13.80, 0.75)$. The $95\%$ confidence interval of the estimated age is $\left(12.4, 15.4\right)$~Gyr.}
    \label{fig:age_distribution}
\end{figure}

The ``fit probability" from the empirical $\chi^2$ distribution was multiplied by the probability found in section \ref{Calibration Star test} to result in a final weight for each of the $1,100$ isochrones. The age distribution is shown in Figure\ \ref{fig:age_distribution}. The weighted average of the age of the $1,100$ isochrones is equal to $13.80$ Gyr and the weighted standard deviation is $0.75$ Gyr. Thus, we measure the absolute age of M92 to be $13.80 \pm 0.75$ Gyr. At 95\% confidence, we find the age to be in the range $12.4-15.4$~Gyr.

\section{Discussion}
\label{sec:discussion}
\subsection{Distance Modulus and Reddening} \label{Distance Modulus and Reddening}
\begin{figure}[htp]
    \centering
    \includegraphics[width=\columnwidth]{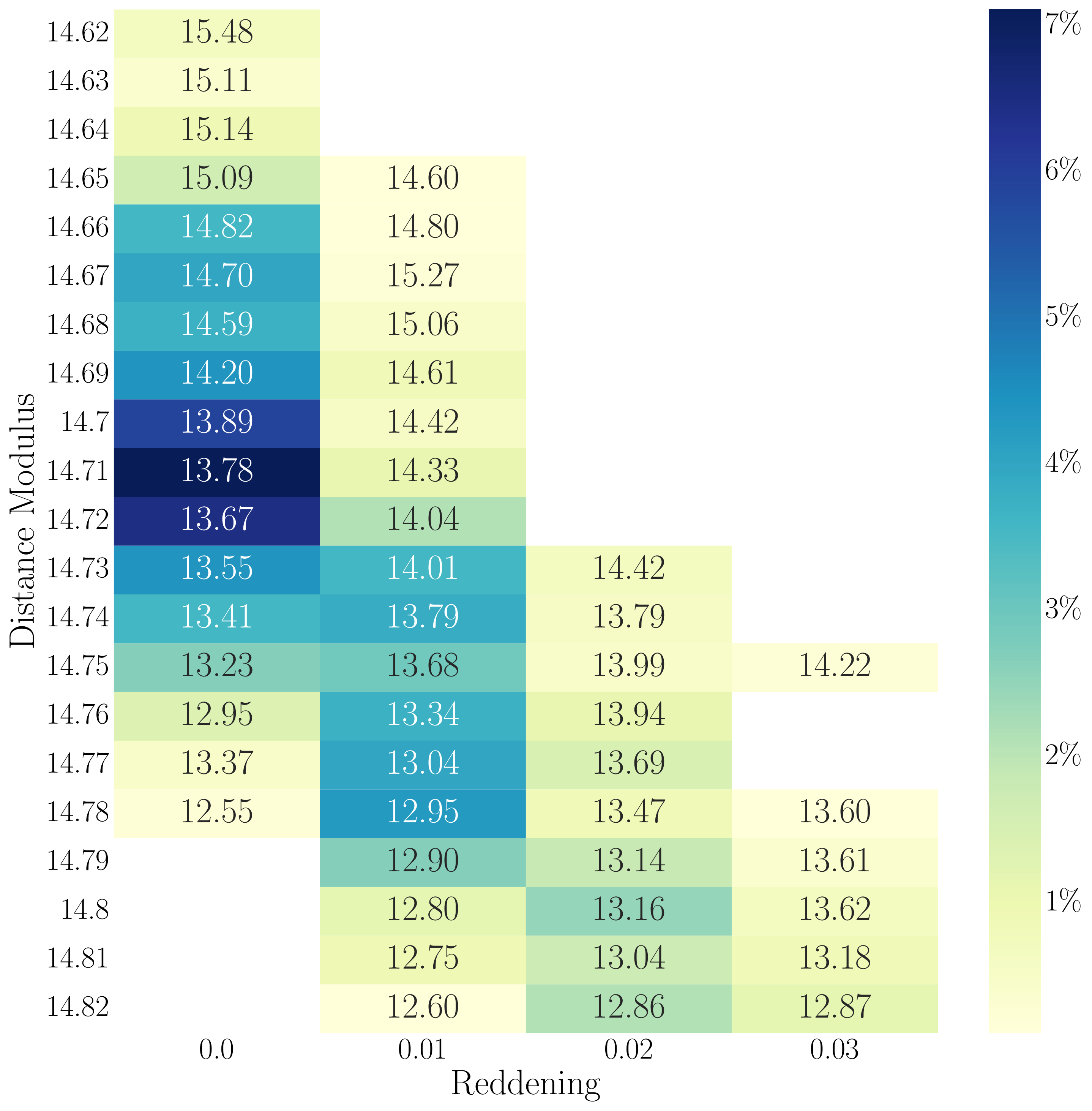}
    \caption{The best-fit age correspond to each combination of distance modulus and reddening. The annotation on each block represents estimated age (Gyr) from each combination. The color represents the occurrence of each combination as a percentage of the total $1,100$ best-fit isochrones. $78$ out of the $1,100$ best-fit isochrones choose distance modulus $= 14.71$ and reddening $= 0.0$ with an estimated age of $13.78$ Gyr.}
    \label{fig:dm_vs_red}
\end{figure}

As described in section \ref{Voronoi Binning}, we tested distance moduli ranging from $14.62$ to $14.82$ (with an increment of $0.01$) and reddening ranging from $0.0$ to $0.05$ (with an increment of $0.01$) for each isochrone. The best fitting age corresponding to each distance modulus and age is shown in Figure \ref{fig:dm_vs_red}.  This figure clearly (and unsurprisingly) indicates that the lower best-fit age favors higher distance modulus. The Pearson correlation coefficient between distance modulus and best-fit age $= -0.780$, indicating a strong negative correlation between  the distance modulus and the best-fit age. This result is expected:  a higher distance modulus will shift the theoretical isochrone in the sCMD in the opposite direction of a shift in the sCMD due to a lower isochrone age and will remain close to the true distribution of stars observed in M92. Figure\ \ref{fig:dm_vs_red} also shows a strong preference toward lower reddening values.

\begin{table*}[]
\caption{\label{tab2}}
\begin{tabular}{llll}
\hline
Source & Age (Gyr) & Distance Modulus $\mu$ (mag)                               & Reddening $E (B - V)$ (mag)\\
\hline
This paper & $13.80 \pm 0.75$ & $14.72 \pm 0.04$ & $0.005 \sim 0.025$ \\
\citet{carretta_distances_2000}& $14.8 \pm 2.5$ &$14.74 \pm 0.07$ & $0.025 \pm 0.005$\\
\citet{vandenberg_models_2002}& $13.5 \pm 1.0$ &$14.62$ & $0.023$\\
\citet{cecco_absolute_2010}& $11.0 \pm 1.5$ &$14.82$ & $0.025 \pm 0.010$\\
\hline
\end{tabular}
\end{table*}

The weighted average of distance modulus is $\mu = 14.72 \pm 0.04$ mag ($D=8.79 \pm 0.16$~kpc), which is the similar to \citet{carretta_distances_2000}, but lower than  \citet{cecco_absolute_2010} and  slightly higher than \citet{baumgardtAccurateDistancesGalactic2021}, who found $\mu = 14.66 \pm 0.04$. Our result has a strong preference towards a low reddening: there are no well-fitting isochrones with $E ( B - V ) = 0.04$ or $0.05$. Since the distribution of reddening is non-symmetric, we select the central $68 \%$ of the distribution and find that the reddening of M92 is in the range $E (B - V) = 0.005 \sim 0.025$ mag, with the distribution skewed to smaller reddening values.  This is within the range of previous results as shown Table~\ref{tab2}

Most of the studies on GC age-dating \citep[e.g.][]{salaris_homogeneous_2002, carretta_distances_2000} rely mainly on main sequence turn-off (MSTO) stars to determine the age of a cluster.  However, we include a wider range of stars with F606W magnitude from $15.925$ to $19.925$. As a result, our study includes not only MSTO stars but also a subset of stars lying on the main sequence (MS) and giant branch (GB). Although MS stars and GB stars are not very sensitive to age, we include them in this study to constrain the distance modulus and reddening of M92. 

To test this idea, we applied the method described in section \ref{Simulated Color-Magnitude Diagram} and section \ref{Isochrone Fitting} to M92 data with only MSTO stars. Due to computational limitations, we generated and fitted $2,000$ sets of sCMDs (1000 of which had been found to be good fits in our previous analysis, and 1000 which were poor fits in our previous analysis). By exclusively using MSTO stars, we were able to determine the age of M92 as  $= 13.88 \pm 0.81$ Gyr with distance modulus $\mu = 14.68 \pm 0.05$ and reddening $E (B - V) = 0.005 \sim 0.045$. These results, which exhibit a slightly higher age and larger uncertainty compared to those in Table~\ref{tab2}, follow expectations: removing MS and GB stars provides more freedom in selecting distance modulus and reddening, thus partially offsetting the impact of the change in isochrone age. In this case, we suggest that the wider range of best-fit reddening values is the cause of the higher uncertainty in age estimated and the slightly lower distance modulus value is likely due to the strong negative correlation between it and the age.

\subsection{Monte Carlo Parameters}
\begin{figure*}[htp]
\centering
\begin{minipage}[b]{.45\textwidth}
\includegraphics[width=0.92\textwidth]{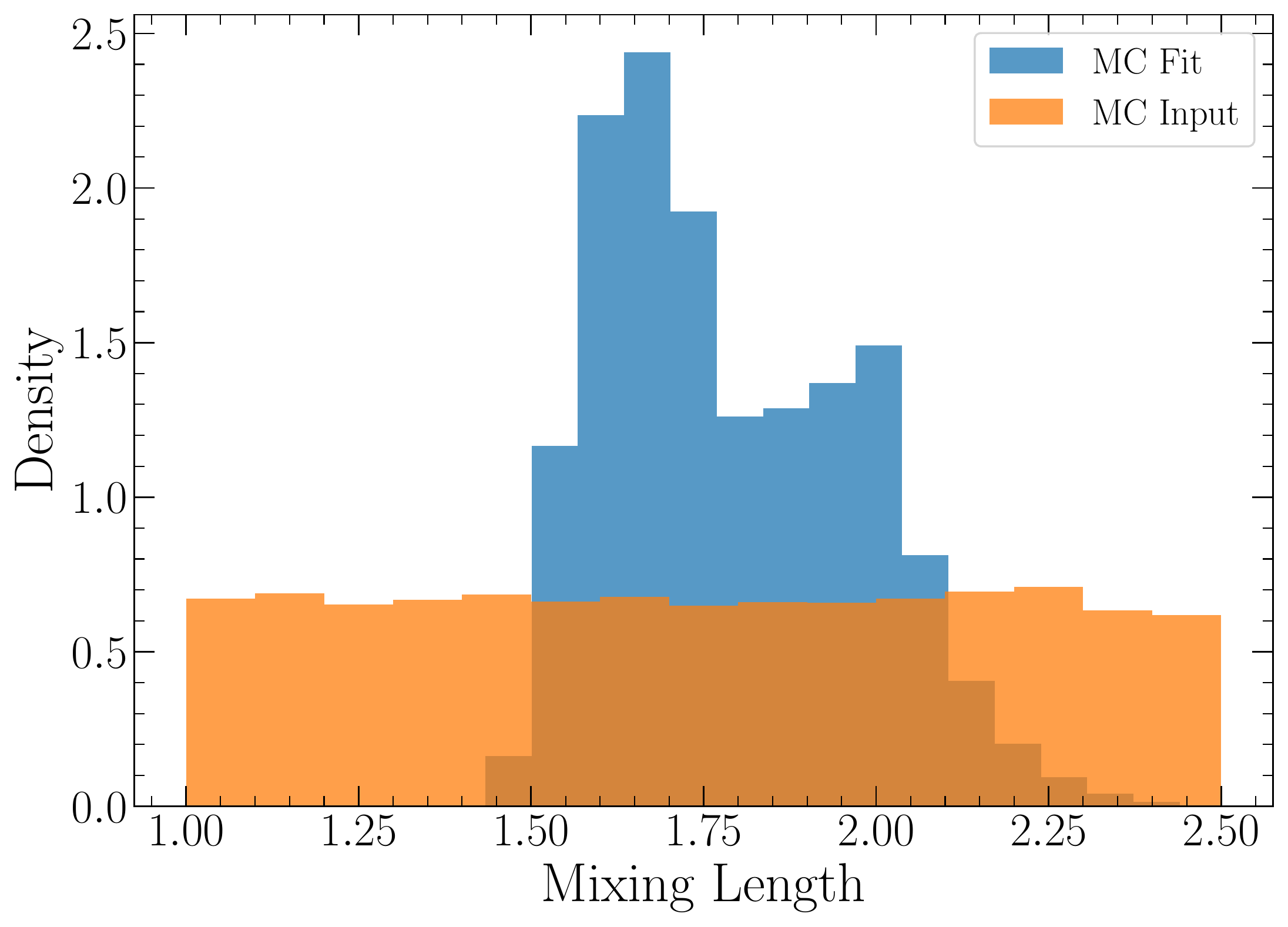}
\caption{The distribution of mixing length as an input parameter (see Table~\ref{tab1}). The solar-calibrated mixing length used as the median set of MC parameters is $1.75$.}\label{fig:cmix_in}
\end{minipage}\qquad
\begin{minipage}[b]{.45\textwidth}
\includegraphics[width=0.9\textwidth]{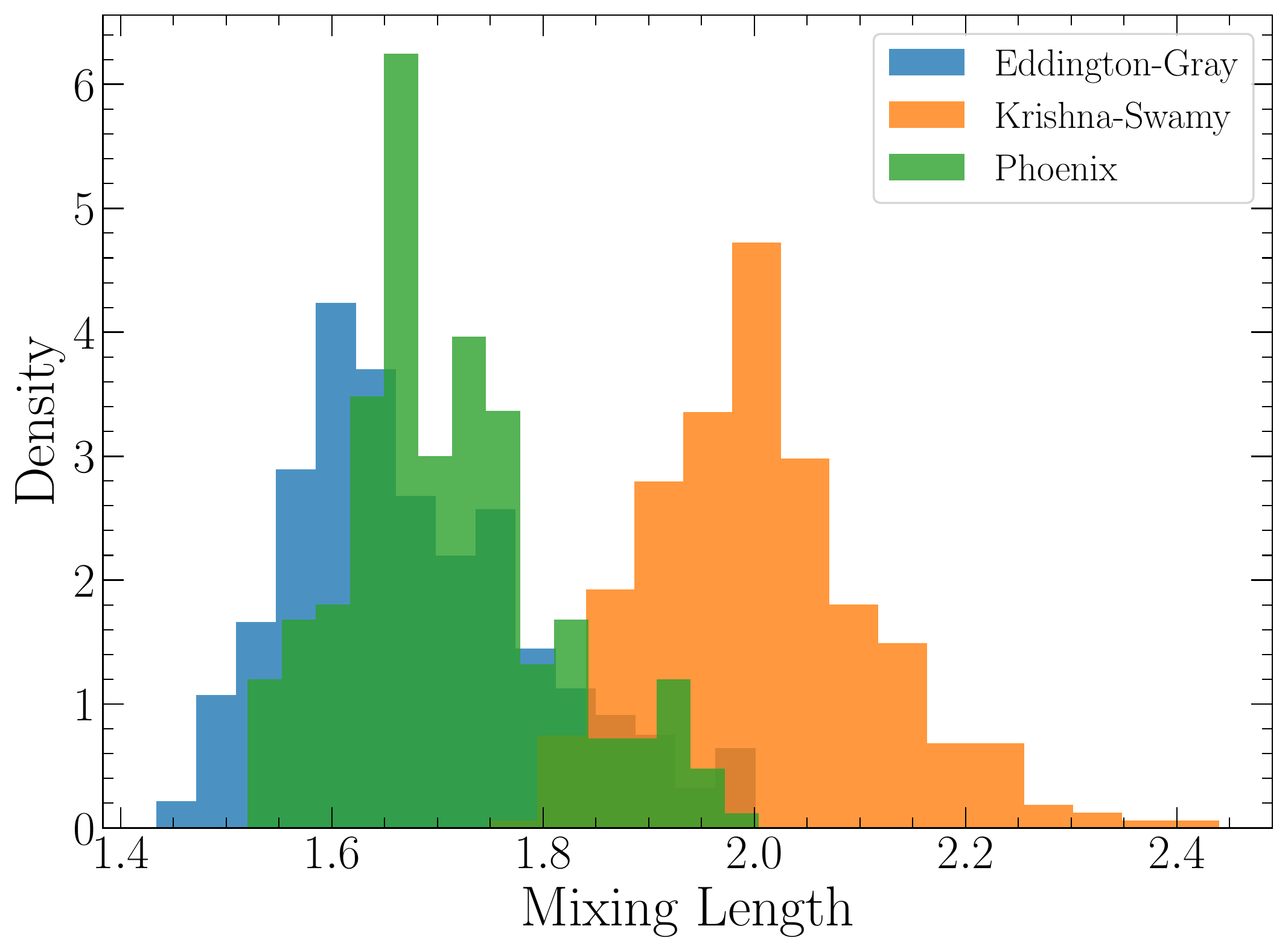}
\caption{The distribution of mixing length from the $1,100$ sets of best-fit isochrones for all three models of atmosphere used. }\label{fig:cmix_boundary}
\end{minipage}
\end{figure*}

\begin{figure*}[htp]
    \centering
    \includegraphics[width=0.8\textwidth]{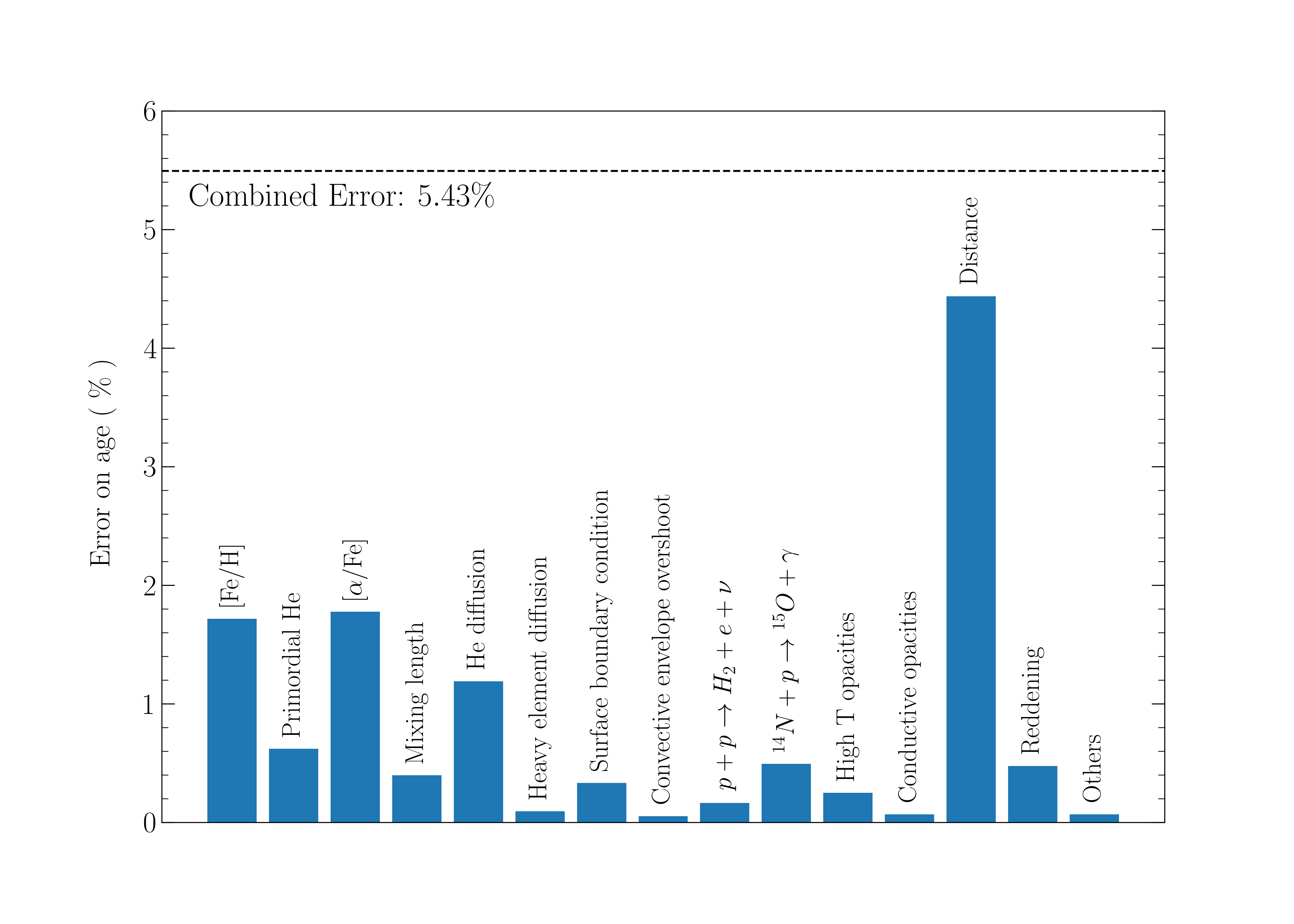}
    \caption{Contributions to the error of the age of M92 from each Monte Carlo parameters with distance modulus and reddening. The covariance and parameters that contribute less than $0.05\%$ (i.e., less than 1\% of the total error budget) are combined as ``others". All the errors are converted to the percentage of the age of M92. The black dotted line represents the combined age error of 5.43\%.}
    \label{fig:error_budget}
\end{figure*}
To determine if the observational data is best fit by a limited range in our MC parameters,  we compare the distribution of input MC parameters  (see Table \ref{tab1}) to the distribution of the MC parameters in the set of  $1,100$  best-fit isochrones. Most parameters have similar distributions,  while differences are found for a few parameters. For example, the distribution of the mixing length is shown in Fig.~\ref{fig:cmix_in}. While a uniform distribution from $1.0$ to $2.5$  was used  as  input, the best-fitting isochrones show a very strong preference for mixing length values between $1.5$ and $2.0$. 

The value of solar-calibrated mixing length depends strongly on the the surface boundary conditions which are used in constructing a stellar model.  The correlation between the mixing length and surface boundary condition for the $1,100$ sets of best-fitting isochrones is shown in Fig.~\ref{fig:cmix_boundary}.  DSEP determines the conditions at the surface of the star using model atmospheres.  The three options used in the MC were PHOENIX model atmospheres (based upon a sophisticated radiative transfer code; see \citealt{Hauschildt1999}) which has a solar calibrated mixing length of 1.7,   the simple Eddington gray model atmosphere which has a solar calibrated mixing length $\alpha_{\textup{MLT}} = 1.7$, and the empirical solar Krishna-Swamy atmosphere which has solar calibrated mixing length $\alpha_{\textup{MLT}} = 2.0$. Both the Gray and Phoenix models of atmosphere prefer lower mixing length value while the Krishna-Swamy model of atmosphere prefers higher mixing length values when fit to M92. The resulted double-peak feature is shown in Fig.~\ref{fig:cmix_in}.  

To determine which MC parameters are most important for in determining the uncertainty in the age estimate for M92, the error budget for each parameter was calculated. For all the MC parameters in Table \ref{tab1}, along with distance modulus and reddening, we performed a maximum likelihood estimation using weighted least square regression method, including the calculation of the covariance matrix. The estimate of the error contributed from each MC parameter is shown in Figure\ \ref{fig:error_budget}. Parameters that contribute less than $0.05\%$ are combined as ``others" and their contribution could be the result of their correlation with other parameters. The distance modulus is the dominant source of error.

 \begin{figure*}[htp]
    \centering
    \includegraphics[width=\textwidth]{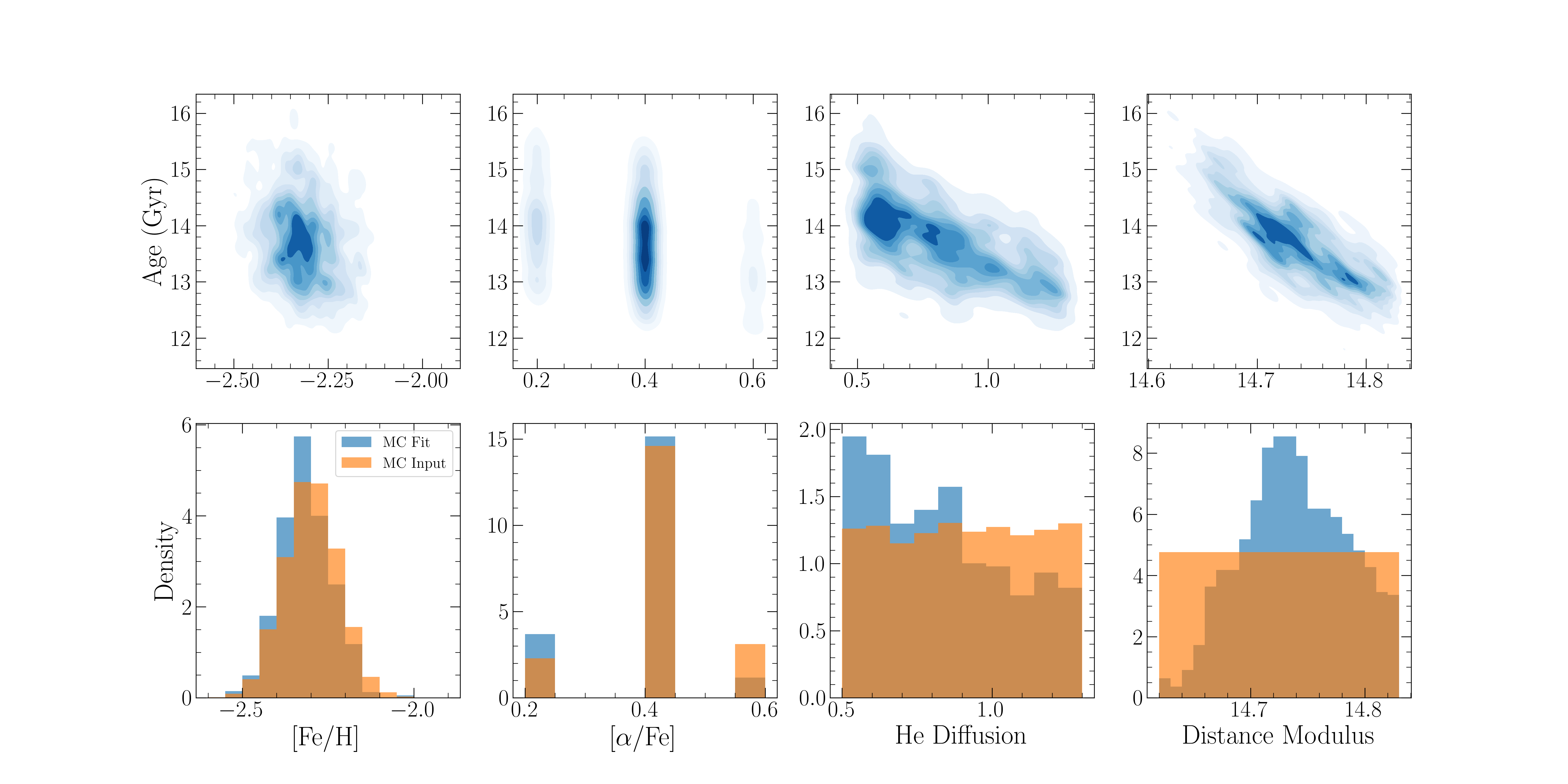}
    \caption{Distribution (bottom panels) of the 4 parameters with $> 1\%$ (or $\gtrsim 140$~Myr) contribution to the age error from Figure~\ref{fig:error_budget} and their correlation with the measured age (top panels). Orange histograms represent the input distribution corresponding to Table\ \ref{tab1} and blue histograms shows the distribution of corresponding MC parameters from the $1,100$ best-fit isochrones.}
    \label{fig:corner}
\end{figure*}

 Since the distance modulus parameter contributes most to the error and might dominate other parameters the value of the distance modulus and reddening was fixed, and the maximum likelihood estimation was repeated  without those two parameters. The result shows a similar contribution from each MC parameters which agree with Figure~\ref{fig:error_budget}. The 4 MC parameters which contribute the most to the error budget were selected and their correlation with estimated age is shown in Figure~\ref{fig:corner}. There is no significant correlation between [Fe/H] and age. This is because M92 has a relatively well determined [Fe/H] value, and since it is very metal-poor, the uncertainty  in the log-scaled [Fe/H] corresponds to a small change in the mass fraction of heavy elements ($Z$). There is a negative correlation between [$\alpha$/Fe] and age, as best-fit isochrones prefer a lower $\alpha$ abundance which leads to a higher estimated age. The helium diffusion coefficient has a MC distribution that is weighted toward a somewhat smaller value than was found in \citet{Thoul1994} and displays an anti-correlation with age.

\section{Conclusion}

We determine the age of M92 using a statistical approach with Monte Carlo simulations which takes into account the uncertainties in the theoretical stellar evolution models and isochrones along with the observed uncertainties in the distance modulus, reddening and composition of M92.  We created $20,000$ sets of Monte Carlo input parameters with 20 variables, which were used to generate $20,000$ sets of theoretical isochrones over an abundance range of $ -2.40 \leq [\textup{Fe/H}] \leq -2.20 $ dex. We use DSEP to construct a set of isochrones from $8$ Gyr to $16$ Gyr with $0.2$ Gyr increment for each set of input parameters. Each isochrone is calibrated using HIP 46120 and HIP 106924, two single, main-sequence stars with accurate colors and absolute magnitudes from HST ACS photometry and Gaia EDR3 parallaxes. 

Each calibrated isochrone is used to generate a sCMD with $4,000,000$ data points. Using the Voronoi Binning method, $800$ bins are generated for each sCMD. HST ACS data for M92 is fit by each set of Voronoi bins with a shift in distance modulus ranging from $14.62$ to $14.82$ with an increment of $0.01$, and reddening ranged from $0.0$ to $0.05$ with an increment of $0.01$. A $\chi^2$ goodness-of-fit parameter was calculated (see eqn.~\ref{eq1}) and compared to the empirical $\chi^2$ distribution generated using HST ACS data combined with artificial star test.

We find that $1,100$ isochrones from the $20,000$ sets of isochrones constructed were within $3\,\sigma$ of the mean of the empirical distribution. The age of M92 is determined by the mean age of the $1,100$ isochrones, weighted by the result from single star calibrations and $\chi^2$ comparison. We find the age of M92 to be $13.80 \pm 0.75$ Gyr, an error of $5.4\%$. The dominant contributor to this uncertainty is the distance modulus, with the metallicity, $\alpha$ enhancement, and treatment of helium diffusion being the other sources of non-negligible error. The fact that the distance to M92, and not stellar physics, dominates the uncertainty points to the importance of precise and accurate distance measurements for further improvements in absolute age measurements. In future papers, we will present absolute age measurements for additional metal-poor GCs and their implications for our understanding of stellar physics and cosmology.

\section*{acknowledgments}
We thank the anonymous referee for a careful review of the paper and helpful comments that improved the presentation of the paper. This material is based upon work supported by the National Science Foundation under Award No.~2007174, by NASA through AR 17043 from the Space Telescope Science Institute (STScI), which is operated by AURA, Inc., under NASA contract NAS5-26555, and  from  The William H. Neukom Institute for Computational Science at Dartmouth College. MBK acknowledges support from NSF CAREER award AST-1752913, NSF grants AST-1910346 and AST-2108962, NASA grant 80NSSC22K0827, and HST-AR-15809, HST-GO-15658, HST-GO-15901, HST-GO-15902, HST-AR-16159, HST-GO-16226, HST-GO-16686, HST-AR-17028, and HST-AR-17043 from STScI.

\software{Dartmouth Stellar Evolution Program \citep{dotter_dartmouth_2008}; Topcat \citep{2005ASPC..347...29T} }

\bibliographystyle{aasjournal}
\bibliography{reference}

\begin{thebibliography}{}
\expandafter\ifx\csname natexlab\endcsname\relax\def\natexlab#1{#1}\fi
\providecommand{\url}[1]{\href{#1}{#1}}
\providecommand{\dodoi}[1]{doi:~\href{http://doi.org/#1}{\nolinkurl{#1}}}
\providecommand{\doeprint}[1]{\href{http://ascl.net/#1}{\nolinkurl{http://ascl.net/#1}}}
\providecommand{\doarXiv}[1]{\href{https://arxiv.org/abs/#1}{\nolinkurl{https://arxiv.org/abs/#1}}}

\bibitem[{{Acharya} {et~al.}(2016){Acharya}, {Carlsson}, {Ekstr{\"o}m},
  {Forss{\'e}n}, \& {Platter}}]{Acharya16}
{Acharya}, B., {Carlsson}, B.~D., {Ekstr{\"o}m}, A., {Forss{\'e}n}, C., \&
  {Platter}, L. 2016, Physics Letters B, 760, 584,
  \dodoi{10.1016/j.physletb.2016.07.032}

\bibitem[{{Adelberger} {et~al.}(2011){Adelberger}, {Garc{\'\i}a}, {Robertson},
  {Snover}, {Balantekin}, {Heeger}, {Ramsey-Musolf}, {Bemmerer}, {Junghans},
  {Bertulani}, {Chen}, {Costantini}, {Prati}, {Couder}, {Uberseder},
  {Wiescher}, {Cyburt}, {Davids}, {Freedman}, {Gai}, {Gazit}, {Gialanella},
  {Imbriani}, {Greife}, {Hass}, {Haxton}, {Itahashi}, {Kubodera}, {Langanke},
  {Leitner}, {Leitner}, {Vetter}, {Winslow}, {Marcucci}, {Motobayashi},
  {Mukhamedzhanov}, {Tribble}, {Nollett}, {Nunes}, {Park}, {Parker},
  {Schiavilla}, {Simpson}, {Spitaleri}, {Strieder}, {Trautvetter}, {Suemmerer},
  \& {Typel}}]{adelberger_solar_2011}
{Adelberger}, E.~G., {Garc{\'\i}a}, A., {Robertson}, R.~G.~H., {et~al.} 2011,
  Reviews of Modern Physics, 83, 195, \dodoi{10.1103/RevModPhys.83.195}

\bibitem[{Anderson {et~al.}(2008)Anderson, Sarajedini, Bedin, King, Piotto,
  Reid, Siegel, Majewski, Paust, Aparicio, Milone, Chaboyer, \&
  Rosenberg}]{anderson_acs_2008}
Anderson, J., Sarajedini, A., Bedin, L.~R., {et~al.} 2008, AJ, 135, 2055,
  \dodoi{10.1088/0004-6256/135/6/2055}

\bibitem[{{Arnold} {et~al.}(2011){Arnold}, {Romanowsky}, {Brodie}, {Chomiuk},
  {Spitler}, {Strader}, {Benson}, \& {Forbes}}]{Arnold11}
{Arnold}, J.~A., {Romanowsky}, A.~J., {Brodie}, J.~P., {et~al.} 2011, \apjl,
  736, L26, \dodoi{10.1088/2041-8205/736/2/L26}

\bibitem[{{Aver} {et~al.}(2015){Aver}, {Olive}, \& {Skillman}}]{Aver2015}
{Aver}, E., {Olive}, K.~A., \& {Skillman}, E.~D. 2015, \jcap, 2015, 011,
  \dodoi{10.1088/1475-7516/2015/07/011}

\bibitem[{Baumgardt \& Vasiliev(2021)}]{baumgardtAccurateDistancesGalactic2021}
Baumgardt, H., \& Vasiliev, E. 2021, MNRAS, 505, 5957,
  \dodoi{10.1093/mnras/stab1474}

\bibitem[{{Bica} \& {Alloin}(1986)}]{Bica86}
{Bica}, E., \& {Alloin}, D. 1986, \aap, 162, 21

\bibitem[{Brown {et~al.}(2014)Brown, Tumlinson, Geha, Simon, Vargas,
  VandenBerg, Kirby, Kalirai, Avila, Gennaro, Ferguson, Muñoz, Guhathakurta,
  \& Renzini}]{brown_quenching_2014}
Brown, T.~M., Tumlinson, J., Geha, M., {et~al.} 2014, ApJ, 796, 91,
  \dodoi{10.1088/0004-637X/796/2/91}

\bibitem[{{Canuto}(1970)}]{Canuto1970}
{Canuto}, V. 1970, \apj, 159, 641, \dodoi{10.1086/150338}

\bibitem[{Cappellari \& Copin(2003)}]{cappellari_adaptive_2003}
Cappellari, M., \& Copin, Y. 2003, MNRAS, 342, 345,
  \dodoi{10.1046/j.1365-8711.2003.06541.x}

\bibitem[{{Carretta} {et~al.}(2009){Carretta}, {Bragaglia}, {Gratton},
  {D'Orazi}, \& {Lucatello}}]{Carretta2009}
{Carretta}, E., {Bragaglia}, A., {Gratton}, R., {D'Orazi}, V., \& {Lucatello},
  S. 2009, \aap, 508, 695, \dodoi{10.1051/0004-6361/200913003}

\bibitem[{Carretta {et~al.}(2000)Carretta, Gratton, Clementini, \&
  Pecci}]{carretta_distances_2000}
Carretta, E., Gratton, R.~G., Clementini, G., \& Pecci, F.~F. 2000, ApJ, 533,
  215, \dodoi{10.1086/308629}

\bibitem[{Cecco {et~al.}(2010)Cecco, Becucci, Bono, Monelli, Stetson,
  Degl’Innocenti, Moroni, Nonino, Weiss, Buonanno, Calamida, Caputo, Corsi,
  Ferraro, Iannicola, Pulone, Romaniello, \& Walker}]{cecco_absolute_2010}
Cecco, A.~D., Becucci, R., Bono, G., {et~al.} 2010, PASP, 122, 991,
  \dodoi{10.1086/656017}

\bibitem[{Chaboyer(1996)}]{chaboyer_age_1996}
Chaboyer, B. 1996, Nuclear Physics B - Proceedings Supplements, 51, 10,
  \dodoi{10.1016/S0920-5632(96)00477-X}

\bibitem[{Chaboyer \& Kim(1995)}]{chaboyerOPALEquationState1995}
Chaboyer, B., \& Kim, Y.-C. 1995, ApJ, 454, 767, \dodoi{10.1086/176529}

\bibitem[{{Chaboyer} {et~al.}(2017){Chaboyer}, {McArthur}, {O'Malley},
  {Benedict}, {Feiden}, {Harrison}, {McWilliam}, {Nelan}, {Patterson}, \&
  {Sarajedini}}]{Chaboyer2017}
{Chaboyer}, B., {McArthur}, B.~E., {O'Malley}, E., {et~al.} 2017, \apj, 835,
  152, \dodoi{10.3847/1538-4357/835/2/152}

\bibitem[{Chakraborty {et~al.}(2015)Chakraborty, deBoer, Mukherjee, \&
  Roy}]{Chakraborty2015}
Chakraborty, S., deBoer, R., Mukherjee, A., \& Roy, S. 2015, Physical Review C,
  91, 045801

\bibitem[{Claret(2004)}]{claretNewGridsStellar2004}
Claret, A. 2004, A \& A, 424, 919, \dodoi{10.1051/0004-6361:20040470}

\bibitem[{{Cohen}(2011)}]{Cohen2011}
{Cohen}, J.~G. 2011, \apjl, 740, L38, \dodoi{10.1088/2041-8205/740/2/L38}

\bibitem[{Collaboration {et~al.}(2021)Collaboration, Brown, Vallenari, Prusti,
  de~Bruijne, Babusiaux, Biermann, Creevey, Evans, Eyer, Hutton, Jansen, Jordi,
  Klioner, Lammers, Lindegren, Luri, Mignard, Panem, Pourbaix, Randich,
  Sartoretti, Soubiran, Walton, Arenou, Bailer-Jones, Bastian, Cropper,
  Drimmel, Katz, Lattanzi, van Leeuwen, Bakker, Cacciari, Castañeda,
  De~Angeli, Ducourant, Fabricius, Fouesneau, Frémat, Guerra, Guerrier,
  Guiraud, Jean-Antoine~Piccolo, Masana, Messineo, Mowlavi, Nicolas,
  Nienartowicz, Pailler, Panuzzo, Riclet, Roux, Seabroke, Sordo, Tanga,
  Thévenin, Gracia-Abril, Portell, Teyssier, Altmann, Andrae, Bellas-Velidis,
  Benson, Berthier, Blomme, Brugaletta, Burgess, Busso, Carry, Cellino, Cheek,
  Clementini, Damerdji, Davidson, Delchambre, Dell'Oro, Fernández-Hernández,
  Galluccio, García-Lario, Garcia-Reinaldos, González-Núñez, Gosset,
  Haigron, Halbwachs, Hambly, Harrison, Hatzidimitriou, Heiter, Hernández,
  Hestroffer, Hodgkin, Holl, Janßen, Jevardat~de Fombelle, Jordan,
  Krone-Martins, Lanzafame, Löffler, Lorca, Manteiga, Marchal, Marrese,
  Moitinho, Mora, Muinonen, Osborne, Pancino, Pauwels, Petit, Recio-Blanco,
  Richards, Riello, Rimoldini, Robin, Roegiers, Rybizki, Sarro, Siopis, Smith,
  Sozzetti, Ulla, Utrilla, van Leeuwen, van Reeven, Abbas, Abreu~Aramburu,
  Accart, Aerts, Aguado, Ajaj, Altavilla, Álvarez, Álvarez Cid-Fuentes,
  Alves, Anderson, Anglada~Varela, Antoja, Audard, Baines, Baker,
  Balaguer-Núñez, Balbinot, Balog, Barache, Barbato, Barros, Barstow,
  Bartolomé, Bassilana, Bauchet, Baudesson-Stella, Becciani, Bellazzini,
  Bernet, Bertone, Bianchi, Blanco-Cuaresma, Boch, Bombrun, Bossini,
  Bouquillon, Bragaglia, Bramante, Breedt, Bressan, Brouillet, Bucciarelli,
  Burlacu, Busonero, Butkevich, Buzzi, Caffau, Cancelliere, Cánovas,
  Cantat-Gaudin, Carballo, Carlucci, Carnerero, Carrasco, Casamiquela,
  Castellani, Castro-Ginard, Castro~Sampol, Chaoul, Charlot, Chemin, Chiavassa,
  Cioni, Comoretto, Cooper, Cornez, Cowell, Crifo, Crosta, Crowley, Dafonte,
  Dapergolas, David, David, de~Laverny, De~Luise, De~March, De~Ridder,
  de~Souza, de~Teodoro, de~Torres, del Peloso, del Pozo, Delbo, Delgado,
  Delgado, Delisle, Di~Matteo, Diakite, Diener, Distefano, Dolding, Eappachen,
  Edvardsson, Enke, Esquej, Fabre, Fabrizio, Faigler, Fedorets, Fernique,
  Fienga, Figueras, Fouron, Fragkoudi, Fraile, Franke, Gai, Garabato,
  Garcia-Gutierrez, García-Torres, Garofalo, Gavras, Gerlach, Geyer, Giacobbe,
  Gilmore, Girona, Giuffrida, Gomel, Gomez, Gonzalez-Santamaria,
  González-Vidal, Granvik, Gutiérrez-Sánchez, Guy, Hauser, Haywood, Helmi,
  Hidalgo, Hilger, Hładczuk, Hobbs, Holland, Huckle, Jasniewicz, Jonker,
  Juaristi~Campillo, Julbe, Karbevska, Kervella, Khanna, Kochoska, Kontizas,
  Kordopatis, Korn, Kostrzewa-Rutkowska, Kruszyńska, Lambert, Lanza, Lasne,
  Le~Campion, Le~Fustec, Lebreton, Lebzelter, Leccia, Leclerc, Lecoeur-Taibi,
  Liao, Licata, Lindstrøm, Lister, Livanou, Lobel, Madrero~Pardo, Managau,
  Mann, Marchant, Marconi, Marcos~Santos, Marinoni, Marocco, Marshall,
  Martin~Polo, Martín-Fleitas, Masip, Massari, Mastrobuono-Battisti, Mazeh,
  McMillan, Messina, Michalik, Millar, Mints, Molina, Molinaro, Molnár,
  Montegriffo, Mor, Morbidelli, Morel, Morris, Mulone, Munoz, Muraveva, Murphy,
  Musella, Noval, Ordénovic, Orrù, Osinde, Pagani, Pagano, Palaversa,
  Palicio, Panahi, Pawlak, Peñalosa~Esteller, Penttilä, Piersimoni, Pineau,
  Plachy, Plum, Poggio, Poretti, Poujoulet, Prša, Pulone, Racero, Ragaini,
  Rainer, Raiteri, Rambaux, Ramos, Ramos-Lerate, Re~Fiorentin, Regibo, Reylé,
  Ripepi, Riva, Rixon, Robichon, Robin, Roelens, Rohrbasser, Romero-Gómez,
  Rowell, Royer, Rybicki, Sadowski, Sagristà~Sellés, Sahlmann, Salgado,
  Salguero, Samaras, Sanchez~Gimenez, Sanna, Santoveña, Sarasso, Schultheis,
  Sciacca, Segol, Segovia, Ségransan, Semeux, Shahaf, Siddiqui, Siebert,
  Siltala, Slezak, Smart, Solano, Solitro, Souami, Souchay, Spagna, Spoto,
  Steele, Steidelmüller, Stephenson, Süveges, Szabados, Szegedi-Elek, Taris,
  Tauran, Taylor, Teixeira, Thuillot, Tonello, Torra, Torra, Turon, Unger,
  Vaillant, van Dillen, Vanel, Vecchiato, Viala, Vicente, Voutsinas, Weiler,
  Wevers, Wyrzykowski, Yoldas, Yvard, Zhao, Zorec, Zucker, Zurbach, \&
  Zwitter}]{collaborationGaiaEarlyData2021}
Collaboration, G., Brown, A. G.~A., Vallenari, A., {et~al.} 2021, A \& A, 649,
  A1, \dodoi{10.1051/0004-6361/202039657}

\bibitem[{deBoer {et~al.}(2014)deBoer, G\"orres, Smith, Uberseder, Wiescher,
  Kontos, Imbriani, Di~Leva, \& Strieder}]{deBoer2014-rates}
deBoer, R.~J., G\"orres, J., Smith, K., {et~al.} 2014, Phys. Rev. C, 90,
  035804, \dodoi{10.1103/PhysRevC.90.035804}

\bibitem[{Demarque {et~al.}(2004)Demarque, Woo, Kim, \&
  Yi}]{demarqueY2IsochronesImproved2004}
Demarque, P., Woo, J.-H., Kim, Y.-C., \& Yi, S.~K. 2004, ApJ Supplement Series,
  155, 667, \dodoi{10.1086/424966}

\bibitem[{Dotter {et~al.}(2008)Dotter, Chaboyer, Jevremović, Kostov, Baron, \&
  Ferguson}]{dotter_dartmouth_2008}
Dotter, A., Chaboyer, B., Jevremović, D., {et~al.} 2008, ApJ Supplement
  Series, 178, 89, \dodoi{10.1086/589654}

\bibitem[{{Ebrahimi} {et~al.}(2020){Ebrahimi}, {Sollima}, {Haghi}, {Baumgardt},
  \& {Hilker}}]{2020MNRAS.494.4226E}
{Ebrahimi}, H., {Sollima}, A., {Haghi}, H., {Baumgardt}, H., \& {Hilker}, M.
  2020, \mnras, 494, 4226, \dodoi{10.1093/mnras/staa969}

\bibitem[{{Eddington}(1926)}]{Eddington1926}
{Eddington}, A.~S. 1926, {The Internal Constitution of the Stars}

\bibitem[{{Ferguson} {et~al.}(2005){Ferguson}, {Alexander}, {Allard}, {Barman},
  {Bodnarik}, {Hauschildt}, {Heffner-Wong}, \& {Tamanai}}]{Ferguson2005}
{Ferguson}, J.~W., {Alexander}, D.~R., {Allard}, F., {et~al.} 2005, \apj, 623,
  585, \dodoi{10.1086/428642}

\bibitem[{{Gaia Collaboration} {et~al.}(2021){Gaia Collaboration}, {Brown},
  {Vallenari}, {Prusti}, {de Bruijne}, {Babusiaux}, {Biermann}, {Creevey},
  {Evans}, {Eyer}, {Hutton}, {Jansen}, {Jordi}, {Klioner}, {Lammers},
  {Lindegren}, {Luri}, {Mignard}, {Panem}, {Pourbaix}, {Randich}, {Sartoretti},
  {Soubiran}, {Walton}, {Arenou}, {Bailer-Jones}, {Bastian}, {Cropper},
  {Drimmel}, {Katz}, {Lattanzi}, {van Leeuwen}, {Bakker}, {Cacciari},
  {Casta{\~n}eda}, {De Angeli}, {Ducourant}, {Fabricius}, {Fouesneau},
  {Fr{\'e}mat}, {Guerra}, {Guerrier}, {Guiraud}, {Jean-Antoine Piccolo},
  {Masana}, {Messineo}, {Mowlavi}, {Nicolas}, {Nienartowicz}, {Pailler},
  {Panuzzo}, {Riclet}, {Roux}, {Seabroke}, {Sordo}, {Tanga}, {Th{\'e}venin},
  {Gracia-Abril}, {Portell}, {Teyssier}, {Altmann}, {Andrae}, {Bellas-Velidis},
  {Benson}, {Berthier}, {Blomme}, {Brugaletta}, {Burgess}, {Busso}, {Carry},
  {Cellino}, {Cheek}, {Clementini}, {Damerdji}, {Davidson}, {Delchambre},
  {Dell'Oro}, {Fern{\'a}ndez-Hern{\'a}ndez}, {Galluccio}, {Garc{\'\i}a-Lario},
  {Garcia-Reinaldos}, {Gonz{\'a}lez-N{\'u}{\~n}ez}, {Gosset}, {Haigron},
  {Halbwachs}, {Hambly}, {Harrison}, {Hatzidimitriou}, {Heiter},
  {Hern{\'a}ndez}, {Hestroffer}, {Hodgkin}, {Holl}, {Jan{\ss}en}, {Jevardat de
  Fombelle}, {Jordan}, {Krone-Martins}, {Lanzafame}, {L{\"o}ffler}, {Lorca},
  {Manteiga}, {Marchal}, {Marrese}, {Moitinho}, {Mora}, {Muinonen}, {Osborne},
  {Pancino}, {Pauwels}, {Petit}, {Recio-Blanco}, {Richards}, {Riello},
  {Rimoldini}, {Robin}, {Roegiers}, {Rybizki}, {Sarro}, {Siopis}, {Smith},
  {Sozzetti}, {Ulla}, {Utrilla}, {van Leeuwen}, {van Reeven}, {Abbas}, {Abreu
  Aramburu}, {Accart}, {Aerts}, {Aguado}, {Ajaj}, {Altavilla}, {{\'A}lvarez},
  {{\'A}lvarez Cid-Fuentes}, {Alves}, {Anderson}, {Anglada Varela}, {Antoja},
  {Audard}, {Baines}, {Baker}, {Balaguer-N{\'u}{\~n}ez}, {Balbinot}, {Balog},
  {Barache}, {Barbato}, {Barros}, {Barstow}, {Bartolom{\'e}}, {Bassilana},
  {Bauchet}, {Baudesson-Stella}, {Becciani}, {Bellazzini}, {Bernet}, {Bertone},
  {Bianchi}, {Blanco-Cuaresma}, {Boch}, {Bombrun}, {Bossini}, {Bouquillon},
  {Bragaglia}, {Bramante}, {Breedt}, {Bressan}, {Brouillet}, {Bucciarelli},
  {Burlacu}, {Busonero}, {Butkevich}, {Buzzi}, {Caffau}, {Cancelliere},
  {C{\'a}novas}, {Cantat-Gaudin}, {Carballo}, {Carlucci}, {Carnerero},
  {Carrasco}, {Casamiquela}, {Castellani}, {Castro-Ginard}, {Castro Sampol},
  {Chaoul}, {Charlot}, {Chemin}, {Chiavassa}, {Cioni}, {Comoretto}, {Cooper},
  {Cornez}, {Cowell}, {Crifo}, {Crosta}, {Crowley}, {Dafonte}, {Dapergolas},
  {David}, {David}, {de Laverny}, {De Luise}, {De March}, {De Ridder}, {de
  Souza}, {de Teodoro}, {de Torres}, {del Peloso}, {del Pozo}, {Delbo},
  {Delgado}, {Delgado}, {Delisle}, {Di Matteo}, {Diakite}, {Diener},
  {Distefano}, {Dolding}, {Eappachen}, {Edvardsson}, {Enke}, {Esquej}, {Fabre},
  {Fabrizio}, {Faigler}, {Fedorets}, {Fernique}, {Fienga}, {Figueras},
  {Fouron}, {Fragkoudi}, {Fraile}, {Franke}, {Gai}, {Garabato},
  {Garcia-Gutierrez}, {Garc{\'\i}a-Torres}, {Garofalo}, {Gavras}, {Gerlach},
  {Geyer}, {Giacobbe}, {Gilmore}, {Girona}, {Giuffrida}, {Gomel}, {Gomez},
  {Gonzalez-Santamaria}, {Gonz{\'a}lez-Vidal}, {Granvik},
  {Guti{\'e}rrez-S{\'a}nchez}, {Guy}, {Hauser}, {Haywood}, {Helmi}, {Hidalgo},
  {Hilger}, {H{\l}adczuk}, {Hobbs}, {Holland}, {Huckle}, {Jasniewicz},
  {Jonker}, {Juaristi Campillo}, {Julbe}, {Karbevska}, {Kervella}, {Khanna},
  {Kochoska}, {Kontizas}, {Kordopatis}, {Korn}, {Kostrzewa-Rutkowska},
  {Kruszy{\'n}ska}, {Lambert}, {Lanza}, {Lasne}, {Le Campion}, {Le Fustec},
  {Lebreton}, {Lebzelter}, {Leccia}, {Leclerc}, {Lecoeur-Taibi}, {Liao},
  {Licata}, {Lindstr{\o}m}, {Lister}, {Livanou}, {Lobel}, {Madrero Pardo},
  {Managau}, {Mann}, {Marchant}, {Marconi}, {Marcos Santos}, {Marinoni},
  {Marocco}, {Marshall}, {Martin Polo}, {Mart{\'\i}n-Fleitas}, {Masip},
  {Massari}, {Mastrobuono-Battisti}, {Mazeh}, {McMillan}, {Messina},
  {Michalik}, {Millar}, {Mints}, {Molina}, {Molinaro}, {Moln{\'a}r},
  {Montegriffo}, {Mor}, {Morbidelli}, {Morel}, {Morris}, {Mulone}, {Munoz},
  {Muraveva}, {Murphy}, {Musella}, {Noval}, {Ord{\'e}novic}, {Orr{\`u}},
  {Osinde}, {Pagani}, {Pagano}, {Palaversa}, {Palicio}, {Panahi}, {Pawlak},
  {Pe{\~n}alosa Esteller}, {Penttil{\"a}}, {Piersimoni}, {Pineau}, {Plachy},
  {Plum}, {Poggio}, {Poretti}, {Poujoulet}, {Pr{\v{s}}a}, {Pulone}, {Racero},
  {Ragaini}, {Rainer}, {Raiteri}, {Rambaux}, {Ramos}, {Ramos-Lerate}, {Re
  Fiorentin}, {Regibo}, {Reyl{\'e}}, {Ripepi}, {Riva}, {Rixon}, {Robichon},
  {Robin}, {Roelens}, {Rohrbasser}, {Romero-G{\'o}mez}, {Rowell}, {Royer},
  {Rybicki}, {Sadowski}, {Sagrist{\`a} Sell{\'e}s}, {Sahlmann}, {Salgado},
  {Salguero}, {Samaras}, {Sanchez Gimenez}, {Sanna}, {Santove{\~n}a},
  {Sarasso}, {Schultheis}, {Sciacca}, {Segol}, {Segovia}, {S{\'e}gransan},
  {Semeux}, {Shahaf}, {Siddiqui}, {Siebert}, {Siltala}, {Slezak}, {Smart},
  {Solano}, {Solitro}, {Souami}, {Souchay}, {Spagna}, {Spoto}, {Steele},
  {Steidelm{\"u}ller}, {Stephenson}, {S{\"u}veges}, {Szabados}, {Szegedi-Elek},
  {Taris}, {Tauran}, {Taylor}, {Teixeira}, {Thuillot}, {Tonello}, {Torra},
  {Torra}, {Turon}, {Unger}, {Vaillant}, {van Dillen}, {Vanel}, {Vecchiato},
  {Viala}, {Vicente}, {Voutsinas}, {Weiler}, {Wevers}, {Wyrzykowski}, {Yoldas},
  {Yvard}, {Zhao}, {Zorec}, {Zucker}, {Zurbach}, \& {Zwitter}}]{EDR3}
{Gaia Collaboration}, {Brown}, A.~G.~A., {Vallenari}, A., {et~al.} 2021, \aap,
  649, A1, \dodoi{10.1051/0004-6361/202039657}

\bibitem[{{Haft} {et~al.}(1994){Haft}, {Raffelt}, \& {Weiss}}]{Haft1994}
{Haft}, M., {Raffelt}, G., \& {Weiss}, A. 1994, \apj, 425, 222,
  \dodoi{10.1086/173978}

\bibitem[{{Hauschildt} {et~al.}(1999){Hauschildt}, {Allard}, \&
  {Baron}}]{Hauschildt1999}
{Hauschildt}, P.~H., {Allard}, F., \& {Baron}, E. 1999, \apj, 512, 377,
  \dodoi{10.1086/306745}

\bibitem[{{Hubbard} \& {Lampe}(1969)}]{Hubbard1969}
{Hubbard}, W.~B., \& {Lampe}, M. 1969, \apjs, 18, 297, \dodoi{10.1086/190192}

\bibitem[{{Iglesias} \& {Rogers}(1996)}]{Iglesias1996}
{Iglesias}, C.~A., \& {Rogers}, F.~J. 1996, \apj, 464, 943,
  \dodoi{10.1086/177381}

\bibitem[{Irwin(2012)}]{irwinFreeEOSEquationState2012}
Irwin, A.~W. 2012, Astrophysics Source Code Library, ascl:1211.002.
\newblock \url{https://ui.adsabs.harvard.edu/abs/2012ascl.soft11002I}

\bibitem[{Joyce \& Chaboyer(2018)}]{joyceNotAllStars2018}
Joyce, M., \& Chaboyer, B. 2018, ApJ, 856, 10, \dodoi{10.3847/1538-4357/aab200}

\bibitem[{Kraft \& Ivans(2003)}]{kraft_globular_2003}
Kraft, R.~P., \& Ivans, I.~I. 2003, PASP, 115, 143, \dodoi{10.1086/345914}

\bibitem[{{Kraft} \& {Ivans}(2003)}]{Kraft2003}
{Kraft}, R.~P., \& {Ivans}, I.~I. 2003, \pasp, 115, 143, \dodoi{10.1086/345914}

\bibitem[{{Krishna Swamy}(1966)}]{KrishnaSwamy1966}
{Krishna Swamy}, K.~S. 1966, \apj, 145, 174, \dodoi{10.1086/148752}

\bibitem[{Lin {et~al.}(2013)Lin, Lucas, \&
  Shmueli}]{linResearchCommentaryToo2013}
Lin, M., Lucas, H.~C., \& Shmueli, G. 2013, Information Systems Research, 24,
  906, \dodoi{10.1287/isre.2013.0480}

\bibitem[{{Lindegren} {et~al.}(2021{\natexlab{a}}){Lindegren}, {Klioner},
  {Hern{\'a}ndez}, {Bombrun}, {Ramos-Lerate}, {Steidelm{\"u}ller}, {Bastian},
  {Biermann}, {de Torres}, {Gerlach}, {Geyer}, {Hilger}, {Hobbs}, {Lammers},
  {McMillan}, {Stephenson}, {Casta{\~n}eda}, {Davidson}, {Fabricius},
  {Gracia-Abril}, {Portell}, {Rowell}, {Teyssier}, {Torra}, {Bartolom{\'e}},
  {Clotet}, {Garralda}, {Gonz{\'a}lez-Vidal}, {Torra}, {Abbas}, {Altmann},
  {Anglada Varela}, {Balaguer-N{\'u}{\~n}ez}, {Balog}, {Barache}, {Becciani},
  {Bernet}, {Bertone}, {Bianchi}, {Bouquillon}, {Brown}, {Bucciarelli},
  {Busonero}, {Butkevich}, {Buzzi}, {Cancelliere}, {Carlucci}, {Charlot},
  {Cioni}, {Crosta}, {Crowley}, {del Peloso}, {del Pozo}, {Drimmel}, {Esquej},
  {Fienga}, {Fraile}, {Gai}, {Garcia-Reinaldos}, {Guerra}, {Hambly}, {Hauser},
  {Jan{\ss}en}, {Jordan}, {Kostrzewa-Rutkowska}, {Lattanzi}, {Liao}, {Licata},
  {Lister}, {L{\"o}ffler}, {Marchant}, {Masip}, {Mignard}, {Mints}, {Molina},
  {Mora}, {Morbidelli}, {Murphy}, {Pagani}, {Panuzzo}, {Pe{\~n}alosa Esteller},
  {Poggio}, {Re Fiorentin}, {Riva}, {Sagrist{\`a} Sell{\'e}s}, {Sanchez
  Gimenez}, {Sarasso}, {Sciacca}, {Siddiqui}, {Smart}, {Souami}, {Spagna},
  {Steele}, {Taris}, {Utrilla}, {van Reeven}, \& {Vecchiato}}]{EDR3pi}
{Lindegren}, L., {Klioner}, S.~A., {Hern{\'a}ndez}, J., {et~al.}
  2021{\natexlab{a}}, \aap, 649, A2, \dodoi{10.1051/0004-6361/202039709}

\bibitem[{{Lindegren} {et~al.}(2021{\natexlab{b}}){Lindegren}, {Bastian},
  {Biermann}, {Bombrun}, {de Torres}, {Gerlach}, {Geyer}, {Hern{\'a}ndez},
  {Hilger}, {Hobbs}, {Klioner}, {Lammers}, {McMillan}, {Ramos-Lerate},
  {Steidelm{\"u}ller}, {Stephenson}, \& {van Leeuwen}}]{EDR3bias}
{Lindegren}, L., {Bastian}, U., {Biermann}, M., {et~al.} 2021{\natexlab{b}},
  \aap, 649, A4, \dodoi{10.1051/0004-6361/202039653}

\bibitem[{{Marcucci} {et~al.}(2013){Marcucci}, {Schiavilla}, \&
  {Viviani}}]{Marcucci13}
{Marcucci}, L.~E., {Schiavilla}, R., \& {Viviani}, M. 2013, \prl, 110, 192503,
  \dodoi{10.1103/PhysRevLett.110.192503}

\bibitem[{{Mar{\'\i}n-Franch} {et~al.}(2009){Mar{\'\i}n-Franch}, {Aparicio},
  {Piotto}, {Rosenberg}, {Chaboyer}, {Sarajedini}, {Siegel}, {Anderson},
  {Bedin}, {Dotter}, {Hempel}, {King}, {Majewski}, {Milone}, {Paust}, \&
  {Reid}}]{2009ApJ...694.1498M}
{Mar{\'\i}n-Franch}, A., {Aparicio}, A., {Piotto}, G., {et~al.} 2009, \apj,
  694, 1498, \dodoi{10.1088/0004-637X/694/2/1498}

\bibitem[{{Marta} {et~al.}(2011){Marta}, {Formicola}, {Bemmerer}, {Broggini},
  {Caciolli}, {Corvisiero}, {Costantini}, {Elekes}, {F{\"u}l{\"o}p}, {Gervino},
  {Guglielmetti}, {Gustavino}, {Gy{\"u}rky}, {Imbriani}, {Junker}, {Lemut},
  {Limata}, {Mazzocchi}, {Menegazzo}, {Prati}, {Roca}, {Rolfs}, {Rossi
  Alvarez}, {Somorjai}, {Straniero}, {Strieder}, {Terrasi}, {Trautvetter}, \&
  {Vomiero}}]{Marta11}
{Marta}, M., {Formicola}, A., {Bemmerer}, D., {et~al.} 2011, \prc, 83, 045804,
  \dodoi{10.1103/PhysRevC.83.045804}

\bibitem[{{Milone} {et~al.}(2012){Milone}, {Piotto}, {Bedin}, {Aparicio},
  {Anderson}, {Sarajedini}, {Marino}, {Moretti}, {Davies}, {Chaboyer},
  {Dotter}, {Hempel}, {Mar{\'\i}n-Franch}, {Majewski}, {Paust}, {Reid},
  {Rosenberg}, \& {Siegel}}]{milone_acs_2012}
{Milone}, A.~P., {Piotto}, G., {Bedin}, L.~R., {et~al.} 2012, \aap, 540, A16,
  \dodoi{10.1051/0004-6361/201016384}

\bibitem[{Milone {et~al.}(2017)Milone, Piotto, Renzini, Marino, Bedin,
  Vesperini, D'Antona, Nardiello, Anderson, King, Yong, Bellini, Aparicio,
  Barbuy, Brown, Cassisi, Ortolani, Salaris, Sarajedini, \& van~der
  Marel}]{miloneHubbleSpaceTelescope2017}
Milone, A.~P., Piotto, G., Renzini, A., {et~al.} 2017, MNRAS, 464, 3636,
  \dodoi{10.1093/mnras/stw2531}

\bibitem[{{Mowla} {et~al.}(2022){Mowla}, {Iyer}, {Desprez},
  {Estrada-Carpenter}, {Martis}, {Noirot}, {Sarrouh}, {Strait}, {Asada},
  {Abraham}, {Brammer}, {Sawicki}, {Willott}, {Bradac}, {Doyon}, {Muzzin},
  {Pacifici}, {Ravindranath}, \& {Zabl}}]{Mowla22}
{Mowla}, L., {Iyer}, K.~G., {Desprez}, G., {et~al.} 2022, \apjl, 937, L35,
  \dodoi{10.3847/2041-8213/ac90ca}

\bibitem[{{Mowlavi} {et~al.}(2012){Mowlavi}, {Eggenberger}, {Meynet},
  {Ekstr{\"o}m}, {Georgy}, {Maeder}, {Charbonnel}, \&
  {Eyer}}]{mowlaviStellarMassAge2012}
{Mowlavi}, N., {Eggenberger}, P., {Meynet}, G., {et~al.} 2012, \aap, 541, A41,
  \dodoi{10.1051/0004-6361/201117749}

\bibitem[{Mészáros {et~al.}(2015)Mészáros, Martell, Shetrone, Lucatello,
  Troup, Bovy, Cunha, García-Hernández, Overbeek, Allende~Prieto, Beers,
  Frinchaboy, García~Pérez, Hearty, Holtzman, Majewski, Nidever, Schiavon,
  Schneider, Sobeck, Smith, Zamora, \&
  Zasowski}]{meszarosExploringAnticorrelationsLight2015}
Mészáros, S., Martell, S.~L., Shetrone, M., {et~al.} 2015, AJ, 149, 153,
  \dodoi{10.1088/0004-6256/149/5/153}

\bibitem[{{O'Malley} {et~al.}(2017){O'Malley}, {McWilliam}, {Chaboyer}, \&
  {Thompson}}]{omalley2017}
{O'Malley}, E.~M., {McWilliam}, A., {Chaboyer}, B., \& {Thompson}, I. 2017,
  \apj, 838, 90, \dodoi{10.3847/1538-4357/aa62a2}

\bibitem[{Paust {et~al.}(2010)Paust, Reid, Piotto, Aparicio, Anderson,
  Sarajedini, Bedin, Chaboyer, Dotter, Hempel, Majewski, Marín-Franch, Milone,
  Rosenberg, \& Siegel}]{paust_acs_2010}
Paust, N. E.~Q., Reid, I.~N., Piotto, G., {et~al.} 2010, AJ, 139, 476,
  \dodoi{10.1088/0004-6256/139/2/476}

\bibitem[{Pietrinferni {et~al.}(2004)Pietrinferni, Cassisi, Salaris, \&
  Castelli}]{pietrinferniLargeStellarEvolution2004}
Pietrinferni, A., Cassisi, S., Salaris, M., \& Castelli, F. 2004, ApJ, 612,
  168, \dodoi{10.1086/422498}

\bibitem[{{Riess} {et~al.}(2021){Riess}, {Casertano}, {Yuan}, {Bowers},
  {Macri}, {Zinn}, \& {Scolnic}}]{riess2021}
{Riess}, A.~G., {Casertano}, S., {Yuan}, W., {et~al.} 2021, \apjl, 908, L6,
  \dodoi{10.3847/2041-8213/abdbaf}

\bibitem[{{Roederer} \& {Sneden}(2011)}]{Roederer2011}
{Roederer}, I.~U., \& {Sneden}, C. 2011, \aj, 142, 22,
  \dodoi{10.1088/0004-6256/142/1/22}

\bibitem[{Salaris \& Weiss(2002)}]{salaris_homogeneous_2002}
Salaris, M., \& Weiss, A. 2002, A \& A, 388, 492,
  \dodoi{10.1051/0004-6361:20020554}

\bibitem[{Sarajedini {et~al.}(2007)Sarajedini, Bedin, Chaboyer, Dotter, Siegel,
  Anderson, Aparicio, King, Majewski, Marín-Franch, Piotto, Reid, \&
  Rosenberg}]{sarajedini_acs_2007}
Sarajedini, A., Bedin, L.~R., Chaboyer, B., {et~al.} 2007, AJ, 133, 1658,
  \dodoi{10.1086/511979}

\bibitem[{{Taylor}(2005)}]{2005ASPC..347...29T}
{Taylor}, M.~B. 2005, in Astronomical Society of the Pacific Conference Series,
  Vol. 347, Astronomical Data Analysis Software and Systems XIV, ed.
  P.~{Shopbell}, M.~{Britton}, \& R.~{Ebert}, 29

\bibitem[{{Thoul} {et~al.}(1994){Thoul}, {Bahcall}, \& {Loeb}}]{Thoul1994}
{Thoul}, A.~A., {Bahcall}, J.~N., \& {Loeb}, A. 1994, \apj, 421, 828,
  \dodoi{10.1086/173695}

\bibitem[{{VandenBerg} {et~al.}(2016){VandenBerg}, {Denissenkov}, \&
  {Catelan}}]{vandenberg2016}
{VandenBerg}, D.~A., {Denissenkov}, P.~A., \& {Catelan}, M. 2016, \apj, 827, 2,
  \dodoi{10.3847/0004-637X/827/1/2}

\bibitem[{VandenBerg {et~al.}(2002)VandenBerg, Richard, Michaud, \&
  Richer}]{vandenberg_models_2002}
VandenBerg, D.~A., Richard, O., Michaud, G., \& Richer, J. 2002, ApJ, 571, 487,
  \dodoi{10.1086/339895}

\bibitem[{{Weisz} {et~al.}(2023){Weisz}, {McQuinn}, {Savino}, {Kallivayalil},
  {Anderson}, {Boyer}, {Correnti}, {Geha}, {Dolphin}, {Sandstrom}, {Cole},
  {Williams}, {Skillman}, {Cohen}, {Newman}, {Beaton}, {Bressan}, {Bolatto},
  {Boylan-Kolchin}, {Brooks}, {Bullock}, {Conroy}, {Cooper}, {Dalcanton},
  {Dotter}, {Fritz}, {Garling}, {Gennaro}, {Gilbert}, {Girardi}, {Johnson},
  {Johnson}, {Kalirai}, {Kirby}, {Lang}, {Marigo}, {Richstein}, {Schlafly},
  {Schmidt}, {Tollerud}, {Warfield}, \& {Wetzel}}]{Weisz23}
{Weisz}, D.~R., {McQuinn}, K. B.~W., {Savino}, A., {et~al.} 2023, arXiv
  e-prints, arXiv:2301.04659, \dodoi{10.48550/arXiv.2301.04659}

\bibitem[{{Xu} {et~al.}(2013){Xu}, {Takahashi}, {Goriely}, {Arnould}, {Ohta},
  \& {Utsunomiya}}]{Xu2013}
{Xu}, Y., {Takahashi}, K., {Goriely}, S., {et~al.} 2013, \nphysa, 918, 61,
  \dodoi{10.1016/j.nuclphysa.2013.09.007}

\bibitem[{{Zinn}(2021)}]{zinn2021}
{Zinn}, J.~C. 2021, \aj, 161, 214, \dodoi{10.3847/1538-3881/abe936}

\end{thebibliography}
\end{document}